\documentclass[a4paper,11pt]{article}
\usepackage{jheppub}% http://jhep.sissa.it/jhep/help/JHEP/TeXclass/DOCS/jheppub.sty
    \usepackage{etoolbox}% http://ctan.org/pkg/etoolbox
    \makeatletter
    \patchcmd{\maketitle}{\@fpheader}{}{}{}
    \makeatother
    
    \usepackage[normalem]{ulem}
\usepackage{tocloft}
\usepackage{amsfonts}
\usepackage{amsmath}
\usepackage{amssymb}
\usepackage{float,tikz, extarrows, tikz-cd}
\usetikzlibrary{decorations.pathmorphing,calc}
\tikzset{snake it/.style={decorate, decoration=snake}}
\usepackage{array}
\usepackage{bigints}
\usepackage{textcomp}
\usepackage{bm}
\usepackage{booktabs}
\usepackage{color}
\usepackage{dsfont}
\usepackage{float}
\usepackage{framed}
\usepackage{graphicx}
\usepackage{tcolorbox}
\usepackage{indentfirst}
\usepackage{mathrsfs}
\usepackage{multirow}
\usepackage{pdflscape}
\usepackage{setspace}
\usepackage{titlesec}
\usepackage{wrapfig}
\usepackage{mathtools}
\usepackage[all]{xy}
\usepackage{young}
\usepackage[vcentermath]{youngtab}
\usepackage{relsize}
\usepackage{stackengine}
\usepackage{verbatim}
\usepackage{slashed}
\usepackage[compat=1.0.0]{tikz-feynman}
\usepackage{adjustbox}
\usepackage{subcaption}
\usepackage{caption}
\def\tilde{\widetilde}

\def\bar{\overline}

\def\1{{\mathds 1}}

\DeclareMathAlphabet{\mathbfsf}{OT1}{cmss}{bx}{n}

\newcommand{\beq}{\begin{equation}\begin{aligned}}
\newcommand{\eeq}{\end{aligned}\end{equation}}

\newcommand{\lam}{\lambda}

\newcommand{\ov}{\over}

\newcommand{\bea}{\begin{eqnarray}}
\newcommand{\eea}{\end{eqnarray}}

\newcommand{\beqa}{\begin{eqnarray}}
\newcommand{\eeqa}{\end{eqnarray}}
\newcommand{\beqar}{\begin{eqnarray*}}
\newcommand{\eeqar}{\end{eqnarray*}}

\def\({\left(} \def\){\right)}
\def\[{\left[} \def\]{\right]}

\title{Symmetry breaking at high temperatures in large N gauge theories}

\author[a]{Soumyadeep Chaudhuri} 
\author[a]{\!, Eliezer Rabinovici}

\vspace{8cm}

\affiliation[a]{Racah Institute, The Hebrew University. Jerusalem 9190401, Israel}

\vspace{8cm}

\emailAdd{chaudhurisoumyadeep@gmail.com} 
\emailAdd{eliezer@mail.huji.ac.il}

\abstract{
Considering marginally relevant and relevant deformations of the weakly coupled $(3+1)$-dimensional large $N$ conformal gauge theories introduced in \cite{Chaudhuri:2020xxb}, we  study the patterns of phase transitions in these systems that lead to a symmetry-broken phase in the high temperature limit. These deformations involve only the scalar fields in the models. The marginally relevant deformations are obtained by varying certain double trace quartic couplings between the scalar fields. The relevant deformations, on the other hand, are obtained by adding masses to the scalar fields while keeping all the couplings frozen at their fixed point values. At the $N\rightarrow\infty$ limit, the RG flows triggered by these deformations approach the aforementioned weakly coupled CFTs in the UV regime. These UV fixed points lie on a conformal manifold with the shape of a circle in the space of couplings.  As shown in \cite{Chaudhuri:2020xxb},  in certain parameter regimes a subset of points on this manifold exhibits thermal order characterized by the spontaneous breaking of a global  $\mathbb Z_2$ or $U(1)$ symmetry  and Higgsing of a subset of gauge bosons at all nonzero temperatures.   We show that the RG flows triggered by the marginally relevant  deformations lead to a weakly coupled IR fixed point which lacks the thermal order. Thus, the systems defined by these RG flows undergo a  transition from a disordered phase at low temperatures to an ordered phase at high temperatures. This provides examples of both inverse symmetry breaking and symmetry nonrestoration.  For the relevant deformations, we demonstrate that a variety of phase transitions are possible depending on the signs and magnitudes of the squares of the masses added to the scalar fields.  Using thermal perturbation theory, we derive the approximate values of the critical temperatures for all these phase transitions.   All the results are obtained at the $N\rightarrow\infty$ limit. Most of them are found in a reliable weak coupling regime and for others we present qualitative arguments. 
}

\begin{document}
\maketitle

\section{Introduction}\label{sec:intro}
Spontaneous breaking of symmetries plays an important role in distinguishing the different  phases of matter.  Usually symmetries that are broken at low temperatures are eventually restored as the temperature becomes sufficiently high.\footnote{Here, as well as in the rest of the paper, we are referring to the spontaneous breaking of ordinary (0-form) global symmetries. There are known instances of spontaneous breaking of higher form symmetries at  high temperatures. One familiar example of this is the spontaneous breaking of the 1-form $\mathbb Z_N$ center symmetry in  pure Yang-Mills theory with $SU(N)$ gauge group. }
 This is true also for  systems in which a symmetry that is unbroken at low temperatures is spontaneously broken at a higher critical temperature  \cite{PhysRevE.72.046107} - a phenomenon called {\it inverse symmetry breaking}.  In each such instance, the broken symmetry  is  restored at an even higher temperature. The ubiquity of such symmetry restoration in nature raised the question of whether there can be models where some symmetry remains broken up to arbitrarily high temperatures.

This question has been explored for several decades in varied contexts ; it has important implications for several areas of physics. We refer the reader to \cite{Weinberg:1974hy, Orloff:1996yn, Bimonte:1999tw, Pinto:1999pg, Mohapatra:1979qt, Langacker:1980kd, Dodelson:1989ii,Dodelson:1991iv,Dvali:1995cj,Meade:2018saz,  Bai:2021hfb, Carena:2021onl} for a sample of the literature. The main difficulty of coming up with a conclusive answer lay in the fact that the theories considered were UV-incomplete, thereby putting  an upper cutoff on the temperatures that can be explored.\footnote{
Examples of symmetry nonrestoration have  been found in certain models with imaginary or random chemical potentials \cite{Komargodski:2017dmc, Aitken:2017ayq, Tanizaki:2017qhf,Dunne:2018hog,Wan:2019oax, Hong:2000rk}. The dynamics of these models suffer from the lack of unitarity. Some holographic models were also explored as candidates for theories with symmetry nonrestoration \cite{Buchel:2009ge,  Donos:2011ut, Gursoy:2018umf, Buchel:2018bzp, Buchel:2020thm, Buchel:2020xdk, Buchel:2020jfs}.  Unlike the phases discussed in this paper, the symmetry-broken phases in these models do not correspond to stable vacua in thermal states.
} 
To overcome this problem certain Wilson-Fisher-like conformal field theories (CFTs) with global $O(N)$  symmetries were studied for both infinite and finite $N$  in fractional dimensions \cite{Chai:2020zgq, Chai:2020onq}.\footnote{There may be issues with unitarity of such Wilson-Fisher-like fixed points at finite $N$ due to the potential existence of operators with complex scaling dimensions \cite{Hogervorst:2015akt}.} The advantage of studying CFTs is that the absence of any intrinsic scale in the theory leads the system to be in the same phase (whatever it may be) at all nonzero temperatures. A symmetry-broken phase was indeed found to occur in the models considered in \cite{Chai:2020zgq, Chai:2020onq}.  This was followed by the construction of (3+1)-dimensional large $N$ conformal gauge theories in \cite{Chaudhuri:2020xxb} where certain global $\mathbb Z_2$ or $U(1)$ symmetries were shown to be broken at all nonzero temperatures. It was also shown that this symmetry breaking is accompanied by the Higgsing of a subset of gauge bosons which leads the system to be in a persistent Brout-Englert-Higgs (BEH) phase. In this context, we refer the reader to \cite{Bajc:2021ope} for examples of similar persistent Higgsing at high temperatures in some asymptotically free large N gauge theories.  
Certain asymptotically safe theories \cite{Litim:2014uca} were also considered in \cite{Bajc:2021ope}, but they failed to show a symmetry-broken phase in the high temperature limit.

The study of the above-mentioned CFTs has demonstrated the existence of persistent symmetry breaking in the large $N$ limit. However, this still leaves open the question of how the persistent symmetry breaking characteristics change in the presence of an extra scale. In this work we address this question by considering relevant and marginally relevant deformations of the large $N$ CFTs introduced in \cite{Chaudhuri:2020xxb}. These CFTs are weakly coupled. They lie on a circle in the space of couplings. In appropriate parameter regimes, a subset of points on this fixed circle demonstrates spontaneous breaking of some global $\mathbb{Z}_2$ or $U(1)$ symmetries at any nonzero temperature. We consider two kinds of deformations of these CFTs with thermal order. As we will discuss shortly, these deformations involve only the scalar fields in the model. At the $N\rightarrow\infty$  limit, the RG flows triggered by these deformations end up at the aforementioned CFTs in the UV regime. So, in this limit these systems provide examples of symmetry nonrestoration in theories with nontrivial UV fixed points. Exploring the IR regimes of these flows, we find interesting patterns of phase transitions at nonzero temperatures. Let us briefly describe these deformations and the corresponding phase transitions in the following two paragraphs.

The first class of deformations that we  consider consists of marginally relevant ones involving variations of certain quartic couplings between the scalar fields.  Such marginally relevant deformations may  exist in four dimensional theories only in the presence of non-Abelian gauge interactions \cite{Zee:1973gn}. This is the case here. We show that the RG flows corresponding to these deformations  lead to a non-Gaussian fixed point in the IR.  The theory remains weakly coupled throughout the  flow allowing the use of perturbation theory to study it. As mentioned earlier, under certain conditions, some of the UV fixed points of these flows demonstrate persistent thermal order. On the other hand, the IR fixed point lacks thermal order. From this we  conclude  that there must be  an inverse symmetry breaking phase transition at an intermediate temperature. By a perturbative analysis, we determine the critical temperature corresponding to this phase transition. Above the critical temperature, the system persistently remains in a symmetry-broken phase. 

The second class of  deformations involves adding masses to the scalar fields while keeping all the couplings frozen at the fixed points exhibiting thermal order. We show that at high temperatures, the effects of these masses are insignificant and the system exists in the same phase as the UV fixed point where  a $\mathbb{Z}_2$ (or $U(1)$) global symmetry remains persistently broken. As the temperature is decreased the effects of the renormalized masses of the scalar fields become increasingly pronounced. At sufficiently low temperatures, such renormalized masses can induce phase transitions in the system. We show that  in certain cases such phase transitions can be studied using thermal perturbation theory. We derive estimates of the critical temperatures corresponding to these  phase transitions. We show that a variety of phases can exist slightly below such critical temperatures. The nature of these phases  crucially depend on the magnitudes and signs of the squares of the renormalized masses. Far below the critical temperatures, the perturbative analysis breaks down. So we cannot say anything definite about the phases in this regime.

We emphasize that all the results mentioned above  are derived in the  $N\rightarrow\infty$ limit. Whether the symmetry nonrestoration in these models persists for finite $N$   remains unresolved.\footnote{In this context, we refer the reader to the recent work \cite{1869270} where  symmetry nonrestoration was found to occur even at finite $N$ in some $d$-dimensional nonlocal CFTs with $1<d<4$.} In the conclusion of the paper we comment on the potential problems that may arise for symmetry nonrestoration at finite $N$.

\paragraph{Organization of the paper:}\mbox{}

In section \ref{sec: dbm review}, we review the models that were introduced in \cite{Chaudhuri:2020xxb}. We discuss the features of the planar beta functions of the couplings in these models and the corresponding fixed points with  special emphasis on the fixed points with thermal order.

In section \ref{sec: marginally relevant deformations:phase transitions}, we demonstrate the existence of marginally relevant deformations of the fixed points exhibiting thermal order. We show that the systems defined by these deformations undergo  inverse symmetry breaking phase transitions at nonzero critical temperatures. We derive  estimates of these critical temperatures.

In section \ref{sec: Massive theory}, we add masses to the scalar fields in the models, and study the  phase transitions induced by these masses at nonzero critical temperatures. We show that there are distinct patterns of phase transitions for the different signs and magnitudes of the squares of these masses. We provide  estimates of the critical temperatures corresponding to these phase transitions.

In section \ref{sec: conclusion}, we conclude by summarizing our results and commenting on how finite $N$ corrections can alter the features found in the $N\rightarrow\infty$ limit.

In appendix \ref{app: positivity of quartic terms} we show that the tree-level quartic terms in the  effective potential of the scalar fields are positive-definite throughout the RG flows discussed in section \ref{sec: marginally relevant deformations:phase transitions}.

In appendix \ref{app: 1-loop beta function of masses} we derive the 1-loop beta functions of the masses of the scalar fields that are introduced in section \ref{sec: Massive theory}.

\section{Review of the double bifundamental models}
\label{sec: dbm review}
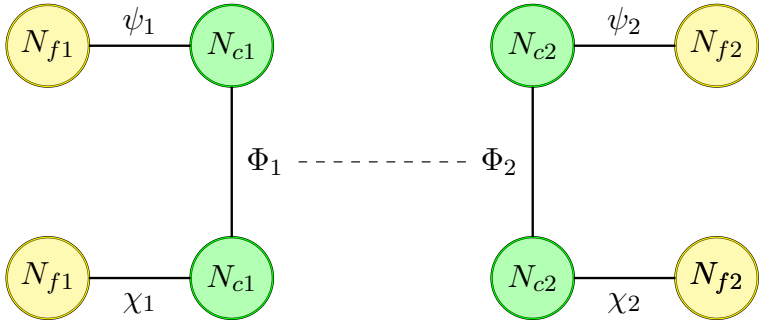
\begin{figure}
\makebox[\textwidth][c]{ \scalebox{1.1}{
\begin{tikzpicture}
    \draw[thick] (-1.8,1.4) circle (14pt) (-1.8,-1.4) circle (14pt)  (1.8,1.4) circle (14pt) (1.8,-1.4) circle (14pt);
    \filldraw[green, fill opacity=0.3] (-1.8,1.4) circle (14pt) (-1.8,-1.4) circle (14pt)  (1.8,1.4) circle (14pt) (1.8,-1.4) circle (14pt);
     \draw[thick]  (-4,1.4) circle (14pt) (-4,-1.4) circle  (14pt)  (4,1.4) circle (14pt) (4,-1.4) circle  (14pt) ;
        \filldraw[yellow, fill opacity=0.3]   (-4,1.4) circle (14pt) (-4,-1.4) circle  (14pt)  (4,1.4) circle (14pt) (4,-1.4) circle  (14pt) ;
    \draw[thick] (-90:14pt) ++(-1.8,1.4)  --($ (90:14pt)+(-1.8,-1.4)$); 
    \draw[thick] (-90:14pt) ++(1.8,1.4)  --($ (90:14pt)+(1.8,-1.4)$); 
 \draw[thick] (180:14pt) ++(-1.8,1.4)  --($ (0:14pt)+(-4,1.4)$); 
\draw[thick] (0:14pt) ++(1.8,1.4)  --($ (180:14pt)+(4,1.4)$); 
\draw[thick] (180:14pt) ++(-1.8,-1.4)  --($ (0:14pt)+(-4,-1.4)$); 
\draw[thick] (0:14pt) ++(1.8,-1.4)  --($ (180:14pt)+(4,-1.4)$); 
\draw[ dashed] (1,0)  --($ (-1,0)$); 
    \node at (-1.8,1.4) {$N_{c1}$};
    \node at (-1.8,-1.4) {$N_{c1}$};
    \node at (1.8,1.4) {$N_{c2}$};
    \node at (1.8,-1.4) {$N_{c2}$};
     \node at (-4,1.4) {$N_{f1}$};
    \node at (-4,-1.4) {$N_{f1}$};
    \node at (4,1.4) {$N_{f2}$};
    \node at (4,-1.4) {$N_{f2}$};
    \node at (4,-1.4) {$N_{f2}$};
     \node at (-1.4,0) {$\Phi_1$};
      \node at (1.4,0) {$\Phi_2$};
       \node at (-2.9,1.7) {$\psi_1$};
        \node at (-2.9,-1.7) {$\chi_1$};
           \node at (2.9,1.7) {$\psi_2$};
        \node at (2.9,-1.7) {$\chi_2$}; 
\end{tikzpicture}}}
\caption{A schematic diagram indicating the representations in which the different fields transform in the double bifundamental models: The two sectors are represented by the two subdiagrams which are connected by a dashed line. This dashed line represents a double trace quartic interaction between the scalar fields in the two sectors. The two green nodes in each sector represent the two $G_i$'s in the  gauge group $G_i\times G_i$ of that sector. Each of these $G_i$'s  is $SO(N_{ci})$ for the RDB model and $SU(N_{ci})$ for the CDB model.  The line connecting the two green nodes in each sector represents the scalar fields in that sector which transform in the bifundamental representation of $G_i\times G_i$. These scalar fields interact  via both single trace and double trace quartic couplings. The two yellow nodes in each sector represent the $N_{fi}$ flavors of the two fermions $\psi_i$ and $\chi_i$ in that sector. Each line connecting a green node and a yellow node indicates that the respective fermion is an $(N_{ci}\times N_{fi})$ matrix which transforms in the fundamental representation of the corresponding $G_i$.
}  \label{fig:dbf}
\end{figure}

In this section we will describe the models that were originally introduced in \cite{Chaudhuri:2020xxb} and which will be the objects of interest in this work. These models have gauge groups of the form 
\beq
G=\prod\limits_{i=1}^2G_i\times G_i
\label{gauge group}
\eeq 
where $G_i$ can be  either $SO(N_{ci})$ or $SU(N_{ci})$.\footnote{The ranks $N_{c1}$ and $N_{c2}$ can be different. } The models where the $G_i$'s are $SO(N_{ci})$ and $SU(N_{ci})$ were called the real  double bifundamental model and the complex double bifundamental model respectively in \cite{Chaudhuri:2020xxb}. These names were coined keeping in mind the representations in which certain scalar fields in the model transform under the gauge group. Henceforth, we will use the abbreviations RDB and CDB for these models. The two sectors of the gauge group are labeled by the index $i$. The matter fields in each sector transform only under the gauge group $(G_i\times G_i)$ in that sector and they are invariant under the gauge transformations in the other sector. The fields in the $i^{\text{th}}$ sector include two sets of massless fermions ($\psi_i$ and $\chi_i$), each of which has $N_{fi}$ flavors and transforms in the fundamental representation of one of the $G_i$'s while being invariant under the other $G_i$. These fermions are Majorana spinors in the RDB model and Dirac spinors in CDB model. In addition to these fermions, there is an $N_{ci}\times N_{ci}$ matrix of massless scalar fields which transform in the bifundamental representation of $G_i\times G_i$. These scalars are real in the RDB model and complex in the CDB model. The scalar fields in each sector interact  via both single trace and double trace quartic couplings. There is an additional double trace interaction coupling the scalar fields in the two sectors. In figure \ref{fig:dbf} we provide a schematic diagram indicating the representations in which the different fields transform in these models.

The renormalized Lagrangians of these models  have the following common form\footnote{In case of the RDB model, $\Phi_i^\dag=\Phi_i^T$ since the scalar fields are real.}:
 \begin{equation}
\begin{split}
\mathcal{L}=&-\frac{1}{2}\sum_{i,\gamma=1}^2\text{Tr}\Big[(F_{i\gamma})_{\mu\nu}(F_{i\gamma})^{\mu\nu}\Big]+i\kappa\sum_{i=1}^2\text{Tr}\Big[\overline{\psi}_i \slashed{D}\psi_i\Big]+i\kappa\sum_{i=1}^2\text{Tr}\Big[\overline{\chi}_i \slashed{D}\chi_i\Big]\\
&+\kappa\sum_{i=1}^2\text{Tr}\Big[\Big(D_{\mu} \Phi_i \Big)^\dag D^{\mu} \Phi_i\Big]-\sum_{i=1}^2\widetilde{h}_i\text{Tr}\Big[(\Phi_i^\dag\Phi_i)^2\Big]-\sum_{i=1}^2\widetilde{f}_i\Big(\text{Tr}\Big[\Phi_i^\dag\Phi_i\Big]\Big)^2\\
&-2\widetilde{\zeta}\text{Tr}\Big[\Phi_1^\dag\Phi_1\Big]\text{Tr}\Big[\Phi_2^\dag\Phi_2\Big],
\end{split}
\label{lagrangian}
\end{equation}
where $\kappa=\frac{1}{2}$ for the RDB model, and $\kappa=1$ for  the CDB model.  The index $\gamma$ distinguishes the two $G_i$'s in the $i^{\text{th}}$ sector. The fermionic fields $\psi_i$ and $\chi_i$ are $(N_{ci}\times N_{fi})$ matrices.

In \cite{Chaudhuri:2020xxb} the fixed points of these models were studied at the Veneziano limit \cite{Veneziano:1976wm} where $N_{ci},N_{fi}\rightarrow\infty$ while $r\equiv\frac{N_{c2}}{N_{c1}}$ and $x_{fi}\equiv\frac{N_{fi}}{N_{ci}}$ are kept finite, and the different couplings scale as
\begin{equation}
g_i^2=\frac{16\pi^2\lambda_i}{N_{ci}},\ \tilde h_i= \frac{16\pi^2 h_i}{N_{ci}},\ \tilde f_i= \frac{16\pi^2 f_i}{N_{ci}^2},\ \tilde \zeta= \frac{16\pi^2 \zeta}{N_{c1}N_{c2}}
\label{dbm: 't Hooft couplings}
\end{equation}
with $g_i$ being the gauge coupling in the $i^{\text{th}}$ sector \footnote{The gauge couplings for the two $G_i$'s in the $i^{\text{th}}$ sector are taken to be equal.}. We will continue to work in this limit. In this limit, there is an orbifold equivalence \cite{Bershadsky:1998cb, Schmaltz:1998bg, Dymarsky:2005nc,Dymarsky:2005uh,Pomoni:2008de, Cherman:2010jj,Hanada:2011ju,Dunne:2016nmc,Aitken:2019shs,Jepsen:2020czw} between the RDB model and the CDB model \cite{Chaudhuri:2020xxb} with the couplings in the two dual theories related by
 \begin{equation}
\begin{split}
\lambda_i^{C}=\frac{\lambda_i^{R}}{2},\ h_i^{C}=2h_i^{R},\ f_i^{C}=2f_i^{R},\ \zeta^{C}=2\zeta^{R}. 
  \end{split}
  \label{duality relations}
\end{equation}
Here the superscript `R' or `C' indicates whether the coupling belongs to the RDB or the CDB model. This planar equivalence between the two models will allow us to restrict our attention to the RDB model. All results that will be derived in this work will have their counterparts in the CDB model. 

Now, the planar beta functions (in the $\overline{\text{MS}}$ scheme) of the different couplings in the RDB model have the following forms:
\begin{equation}
\begin{split}
&\beta_{\lambda_i}=-\Big(\frac{21-4 x_{fi}}{6}\Big)\lambda_i^2+\Big(\frac{-27+13 x_{fi}}{6}\Big)\lambda_i^3+\cdots\  ,\\
&\beta_{h_i}=16 h_i^2-6 h_i\lambda_i+\frac{3}{16}\lambda_i^2+\cdots\ ,\\
&\beta_{f_i}=8 f_i^2+32 f_ih_i-6 f_i\lambda_i+24h_i^2+\frac{9}{16}\lambda_i^2+8\zeta^2+\cdots\ ,\\
&\beta_{\zeta}=\zeta\Bigg[8 f_1+8 f_2+16 h_1+16 h_2-3\lambda_1-3\lambda_2\Bigg]+\cdots\ .
\end{split}
\label{rdb: beta fns}
\end{equation}
The dots indicate higher order corrections. There are several features of these beta functions which will be important for our analysis in this work. We would like to mention two of them here. Firstly, note that these beta functions are independent of the ratio $r=\frac{N_{c2}}{N_{c1}}$. This means that the RG flows that we will study in section \ref{sec: marginally relevant deformations:phase transitions} will be independent of this ratio. Secondly,  the beta functions of the single trace couplings ($\lambda_i$ and $h_i$) are independent of the double trace couplings ($f_i$ and $\zeta$).\footnote{This is a general feature of planar beta functions of single trace couplings in large N gauge theories \cite{Dymarsky:2005uh}.} These properties of the beta functions were shown to survive up to all orders at the Veneziano limit  in \cite{Chaudhuri:2020xxb}.

Based on the forms of the above beta functions, one can show that there are unitary perturbative fixed points\footnote{These fixed points are akin to the Banks-Zaks-Caswell fixed points \cite{Belavin:1974gu, PhysRevLett.33.244, Banks:1981nn} in more familiar QCDs.} which are of the following two kinds.
\begin{enumerate}
\item There is a discrete set of fixed points where the two sectors are decoupled and the couplings have the following values at leading order:
\beq 
&\lam_i={21-4x_{fi}\ov 13 x_{fi}-27} ,\ h_i={3-\sqrt{6}\ov 16}\lam_i,\ f_i={2\sqrt{6} +\sigma_i  \sqrt{18\sqrt{6}-39}\ov 16}\lam_i,\ \zeta=0
\eeq
with $\sigma_i=\pm1$.
Note that to get unitary fixed points in a perturbative regime, one must set $x_{fi}=\frac{21}{4}-\epsilon_i$ with $0<\epsilon_i\ll1$.

\item The second class of fixed points consists of theories where the two sectors are coupled. These fixed points exist only when $x_{f1}=x_{f2}\equiv x_f$. They form a conformal manifold which has the shape of a circle in the space of couplings\footnote{A couple of points on this circle where $\zeta=0$ have already been included in the first class.}. The values of the different couplings on this fixed circle at leading order  are as follows:
\beq
&\lam_1=\lam_2\equiv\lam={21-4x_{f}\ov 13 x_{f}-27} ,\ h_1=h_2\equiv h={3-\sqrt{6}\ov 16}\lam,\\
&f_p=\frac{\sqrt{6}}{8}\lam,\ f_m^2+\zeta^2=\Big({18\sqrt{6}-39\ov 256}\Big)\lam^2.
\label{conformal manifold:1-loop fixed point}
\eeq
where $f_p\equiv\frac{f_1+f_2}{2}$ and $f_m\equiv \frac{f_1-f_2}{2}$.
\end{enumerate} 
All these fixed points survive under higher loop corrections at the planar limit and have stable effective potentials (at least up to leading order in perturbation theory)\cite{Chaudhuri:2020xxb}. Moreover, in appropriate regimes of the ratio $r$ certain points on the fixed circle exhibit thermal order, i.e., a $\mathbb{Z}_2$ symmetry in the  sector with the smaller rank is spontaneously broken at all nonzero temperatures. The relevant  $\mathbb{Z}_2$ symmetry in the $i^{\text{th}}$ sector transforms the different fields in that sector as follows:
\begin{equation}
\begin{split} 
&\Phi_i\rightarrow \mathcal{T}_i\Phi_i,\ \ \psi_i\rightarrow \mathcal{T}_i \psi_i ,\ (V_{i1})_\mu\rightarrow \mathcal{T}_i (V_{i1})_\mu \mathcal{T}_i^{-1},
\end{split}
\label{Z2 baryon symmetry in RDB model}
\end{equation}
where $\mathcal{T}_i$ is the following $N_{ci}\times N_{ci}$ diagonal matrix:
\beq
\mathcal{T}_i\equiv\text{diag}\{-1,1,\cdots,1\}.
\eeq
It leaves the fields in the other sector unchanged. These symmetry transformations map one class of gauge-equivalent field configurations to another.\footnote{See appendix A of \cite{Chaudhuri:2020xxb} for a proof of this.} These global symmetries  were called baryon symmetries in \cite{Chaudhuri:2020xxb} because  the corresponding order parameters are the expectation values of the determinants of the scalar fields.

\begin{figure}
\centering
\includegraphics[scale=0.6]{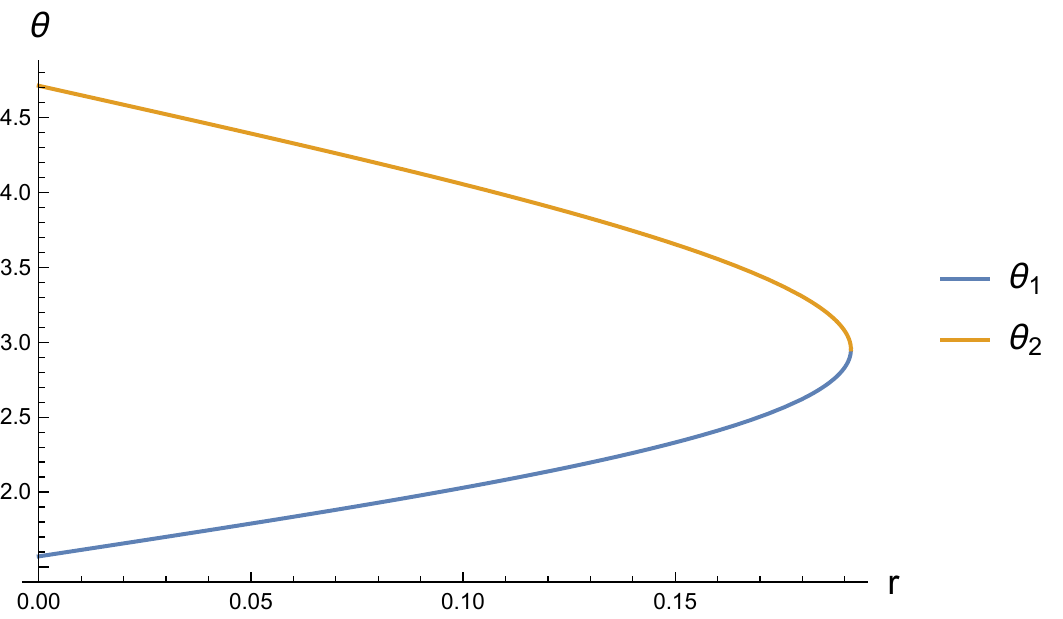}\,
\caption{Graphs of $\theta_1$ and $\theta_2$ against $r$: For each value of $r<r_{\text{max}}\equiv\sqrt{\frac{6\sqrt{6}-13}{61-6\sqrt{6}}}$, $\theta_1$ and $\theta_2$ are the endpoints of the angular interval on the fixed circle which exhibits thermal order in the second sector. They always lie in the regime $\frac{\pi}{2}<\theta_1<\theta_2<\frac{3\pi}{2}$. As $r\rightarrow 0$, $\theta_1\rightarrow\frac{\pi}{2}$ and $\theta_2\rightarrow\frac{3\pi}{2}$. As $r\rightarrow r_{\text{max}}$, $\theta_{1,2}\rightarrow\cos^{-1}\Big(-\frac{\sqrt{61-6\sqrt{6}}}{4\sqrt{3}}\Big)\approx 2.952$.}
\label{theta_r_graph}
\end{figure}

When $N_{c2}<N_{c1}$, the baryon symmetry in the first sector remains unbroken at all temperatures. On the other hand, the baryon symmetry in the second sector  can be spontaneously broken in a thermal state for a subset of  points on the conformal manifold  when $r< r_{\text{max}}\equiv\sqrt{\frac{6\sqrt{6}-13}{61-6\sqrt{6}}}\approx 0.191$. These points can be expressed  simply by defining the following polar coordinates in the $f_m$-$\zeta$ plane:
\beq
f_m=R\sin\theta,\ \zeta=R\cos\theta,
\label{polar coordinates}
\eeq
where $\theta\in[0,2\pi)$.
 For each $r<r_{\text{max}}$, the fixed points exhibiting thermal order lie in the angular interval $\theta\in(\theta_1,\theta_2)$ where\footnote{One can check that the angular interval given here is equivalent to the intervals of $f_m$ and $\zeta$ given in \cite{Chaudhuri:2020xxb}.}
 \beq
 \theta_1=\cos^{-1}(\nu_1),\  \theta_2=\pi-\text{sgn}(r-r_0)\Big(\pi-\cos^{-1}(\nu_2)\Big)
 \label{domain of fixed points with thermal order}
 \eeq
 with $r_0\equiv\frac{\sqrt{18\sqrt{6}-39}}{12}$ and
 \beq
 &\nu_1\equiv r\Bigg[\frac{-12+\sqrt{\Big(18\sqrt{6}-183\Big)r^2+\Big(18\sqrt{6}-39\Big)}}{\Big(1+r^2\Big)\sqrt{18\sqrt{16}-39}}\Bigg]\ ,\\
 &\nu_2\equiv r\Bigg[\frac{-12-\sqrt{\Big(18\sqrt{6}-183\Big)r^2+\Big(18\sqrt{6}-39\Big)}}{\Big(1+r^2\Big)\sqrt{18\sqrt{16}-39}}\Bigg]\ .
 \eeq
Here the range of the function $\cos^{-1}$ is the interval $[0,\pi]$. The plots of $\theta_1$ and $\theta_2$ as functions of $r$ are shown in figure \ref{theta_r_graph}. From these plots, one can see that for all values of $r\in(0,r_{\text{max}})$ the end points of the above angular interval satisfy $\frac{\pi}{2}<\theta_1<\theta_2<\frac{3\pi}{2}$ with $\theta_1\rightarrow\frac{\pi}{2}$ and $\theta_2\rightarrow\frac{3\pi}{2}$ as $r\rightarrow 0$, and  $\theta_{1,2}\rightarrow\cos^{-1}\Big(-\frac{\sqrt{61-6\sqrt{6}}}{4\sqrt{3}}\Big)\approx 2.952$ as $r\rightarrow r_{\text{max}}$. Thus, $\cos\theta$ (or equivalently, $\zeta=R\cos\theta$)  is always negative for the fixed points demonstrating thermal order. 

When $N_{c1}<N_{c2}$, one would get similar fixed  points with spontaneous breaking  of the baryon symmetry in the first sector at all nonzero temperatures. These fixed points can be obtained from those demonstrating thermal order in the second sector by the transformations $r\rightarrow\frac{1}{r}$ and  $\theta\rightarrow 2\pi-\theta$.

It was shown in \cite{Chaudhuri:2020xxb} that for all the fixed points where the baryon symmetry in a sector is spontanously broken at nonzero temperatures, this phenomenon is accompanied by the Higgsing of half of the gauge bosons in the same sector. This leads the system to be in a persistent Brout-Englert-Higgs (BEH) phase at all nonzero temperatures.  

A similar spontaneous breaking of a baryon symmetry at nonzero temperatures  and a persistent BEH phase were also found at the corresponding fixed points of the dual CDB model. In this case, the relevant baryon symmetry in each sector is a $U(1)$ symmetry which can be obtained by replacing $\mathcal{T}_i$ by the $N_{ci}\times N_{ci}$ diagonal matrix defined below:
\beq
(\mathcal{T}_i)_\phi\equiv\text{diag}\{e^{i\phi},1,\cdots,1\}.
\eeq 
with $\phi\in [0,2\pi)$.

An important result which was derived in \cite{Chaudhuri:2020xxb} is that the above-mentioned baryon symmetries  are not broken in a thermal state for the fixed points where the two sectors are decoupled. We will find this result to be  consequential  as  one of these fixed points is the IR limit of the RG flows corresponding to the marginally relevant deformations discussed in section \ref{sec: marginally relevant deformations:phase transitions}.

\section{Phase transitions for marginally relevant deformations}
\label{sec: marginally relevant deformations:phase transitions}
In this section we will consider certain marginally relevant deformations of the points on the  fixed circle in the RDB model  by varying the quartic couplings between the scalars  fields. In subsection \ref{subsec: marginal deformations of fixed points} we will demonstrate the existence of such a marginally relevant deformation for each point on the fixed circle. Later in subsection \ref{subsec: RG flow} we will study the RG flows triggered by  these deformations. We will show that in the IR limit  these RG flows take the corresponding systems to a unique fixed point that lacks thermal order. Therefore, at low temperatures all these systems are in a disordered phase. On the other hand, in the UV limit some of these systems flow to the points on the fixed circle which exhibit thermal order. This means that at high temperatures these systems are in an ordered phase. Therefore, each of these systems must  undergo  an inverse symmetry breaking  phase transition at some critical temperature. Above this critical temperature, the system remains persistently in a symmetry-broken phase. In subsection \ref{subsec: critical temp estimate}, we will provide an estimate of this critical temperature.  
\subsection{Deformations of the double trace quartic couplings}
\label{subsec: marginal deformations of fixed points}
Let us now turn our attention to the deformations of the quartic couplings away from their values on the fixed circle. The  deformations that are of interest to us involve only the double trace couplings ($f_i$ and $\zeta$). Since the beta functions of the single trace couplings ($\lambda_i$ and $h_i$) are independent of the double trace ones at the Veneziano limit, these couplings  remain frozen at their fixed point    values throughout the RG flows triggered by such deformations. Therefore, all such RG flows  take place only in a subspace where $\lambda_1=\lambda_2=\lambda$ and $h_1=h_2=h$ with $\lambda$ and $h$ being the values given in \eqref{conformal manifold:1-loop fixed point} at leading order in perturbation theory. 
In this subspace the full planar beta functions of the double trace couplings have the following forms \footnote{We refer the reader to section 5 of \cite{Chaudhuri:2020xxb} for a derivation of these forms of the  beta functions of the double trace couplings up to all orders in perturbation theory at the planar limit.}:
\begin{equation}
\begin{split}
&\beta_{f_p}=F_1(f_p, \sqrt{f_m^2+\zeta^2}),\ \beta_{f_m}=f_m \tilde F_2(f_p, \sqrt{f_m^2+\zeta^2}),\ \beta_{\zeta}=\zeta \tilde F_2(f_p, \sqrt{f_m^2+\zeta^2}),
\end{split}
\end{equation}
where $F_1$ and $\tilde F_2$ are two functions of $f_p$ and $\sqrt{f_m^2+\zeta^2}$. Switching to the polar coordinates introduced in \eqref{polar coordinates}, the beta functions can be re-expressed as 
\begin{equation}
\begin{split}
&\beta_{f_p}=F_1(f_p, R),\ \beta_{R}= F_2(f_p, R),\ R\beta_{\theta}=0,
\end{split}
\end{equation}
where $F_2(f_p, R)\equiv  R \tilde F_2(f_p, R)$. The expressions of $F_1(f_p,R)$ and $F_2(f_p,R)$  at leading order are 
\begin{equation}
\begin{split}
&F_1(f_p, R)=(8 f_p+32  h-6 \lambda) f_p+8R^2+24h^2+\frac{9}{16}\lambda^2\ ,\\
&F_2(f_p, R)=R(16 f_p+32 h-6\lambda)\ .\\
\end{split}
\label{F1, F2 leading order expressions}
\end{equation}
We can see that $\beta_\theta=0$ for all points in the space of the double trace couplings where $R\neq 0$. Moreover, the beta functions of $f_p$ and $R$ are independent of the angular coordinate $\theta$. Therefore, for a fixed point of these beta functions where $R\neq 0$, a shift in the value of $\theta$  spans the conformal manifold. Another important consequence of the vanishing of $\beta_\theta$ is that the projections of all RG flows on the $f_m$-$\zeta$ plane are radially directed. Furthermore, the fact that $\beta_{f_p}$ and $\beta_{R}$ are independent of $\theta$ ensures that such flows are identical for all values of  $\theta$.

Now, suppose $f_{0p}$ and $R_0$ are the values of $f_p$ and $R$ for the points on the conformal manifold. The leading order expressions of $f_{0p}$ and $R_0$ are (see \eqref{conformal manifold:1-loop fixed point})
\beq
f_{0p}=\frac{\sqrt{6}}{8}\lambda,\ R_0=\Big(\frac{\sqrt{18\sqrt{6}-39}}{16}\Big)\lambda.
\eeq
Consider the following small deformations of $f_p$ and $R$ away from their values on the fixed circle:
\beq
f_p=f_{0p}+\delta f_p,\ R=R_0+\delta R.
\eeq
The beta functions of these deformations in the values of $f_p$ and $R$ (at linear order) are
\begin{equation}
\begin{split}
&\beta_{\delta f_p}=\frac{\partial F_1}{\partial f_p}(f_{0p}, R_0) \delta f_p+\frac{\partial F_1}{\partial R}(f_{0p}, R_0) \delta R,\\
& \beta_{\delta R}=\frac{\partial  F_2}{\partial f_p}(f_{0p}, R_0)\delta f_p+\frac{\partial  F_2}{\partial R}(f_{0p}, R_0)\delta R.
\end{split}
\end{equation}
The coefficient matrix multiplying the deformations in the above equations is
\beq
\mathcal{M}
=\begin{pmatrix}
\frac{\partial F_1}{\partial f_p}(f_{0p}, R_0) & \frac{\partial F_1}{\partial R}(f_{0p}, R_0)\\
\frac{\partial  F_2}{\partial f_p}(f_{0p}, R_0) & \frac{\partial  F_2}{\partial R}(f_{0p}, R_0)
\end{pmatrix}
=16R_0 
 \begin{pmatrix}
0 & 1\\
1 & 0
\end{pmatrix}.
\eeq
The second equality in the above equation is obtained by   substituting $F_1$, $F_2$ and the couplings $h$, $f_{0p}$ and $R_0$ by their leading order expressions. 
The eigenvalues of this matrix  and the corresponding eigenvectors are as follows:
\begin{equation}
\begin{split}
& v_1=-16R_0,\ v_2=16 R_0,\\  
& e_1=\begin{pmatrix}1 \\ -1\end{pmatrix},\ e_2=\begin{pmatrix}1 \\ 1\end{pmatrix}.
\end{split}
 \end{equation}
 Therefore, the deformations along $e_1$ and $e_2$ are marginally relevant and marginally irrelevant respectively. The existence of a marginally relevant deformation clearly demonstrates that all points on the conformal manifold are UV fixed points of certain RG flows. Next, let us study the features of these RG flows. 

\subsection{RG flows triggered by the marginally relevant deformations}
\label{subsec: RG flow}
To analyze the RG flows triggered by the marginally relevant deformations, we will rely on the 1-loop expressions of the beta functions. This approximation is justified near the UV fixed points on the conformal manifold as these fixed points are weakly coupled. However, the RG flows triggered by the marginally relevant deformations can take the system to a strongly coupled regime. We will show that this is not the case when we choose the marginally relevant deformation such that it has a radially inward component in the $f_m$-$\zeta$ plane, i.e., $\delta R <0$. Therefore, in this case one can trust the results based on the 1-loop beta functions. If the deformation is chosen to be in the opposite direction, i.e., $\delta R>0$, then the theory  indeed flows to a strongly coupled regime. This can be verified by noticing that the RG flows obtained from the 1-loop beta functions  lead to divergences of the couplings at finite energy scales.

\begin{figure}[H]
\begin{subfigure}{.5\textwidth}
  \centering
  \scalebox{1}{\includegraphics[width=.8\linewidth]{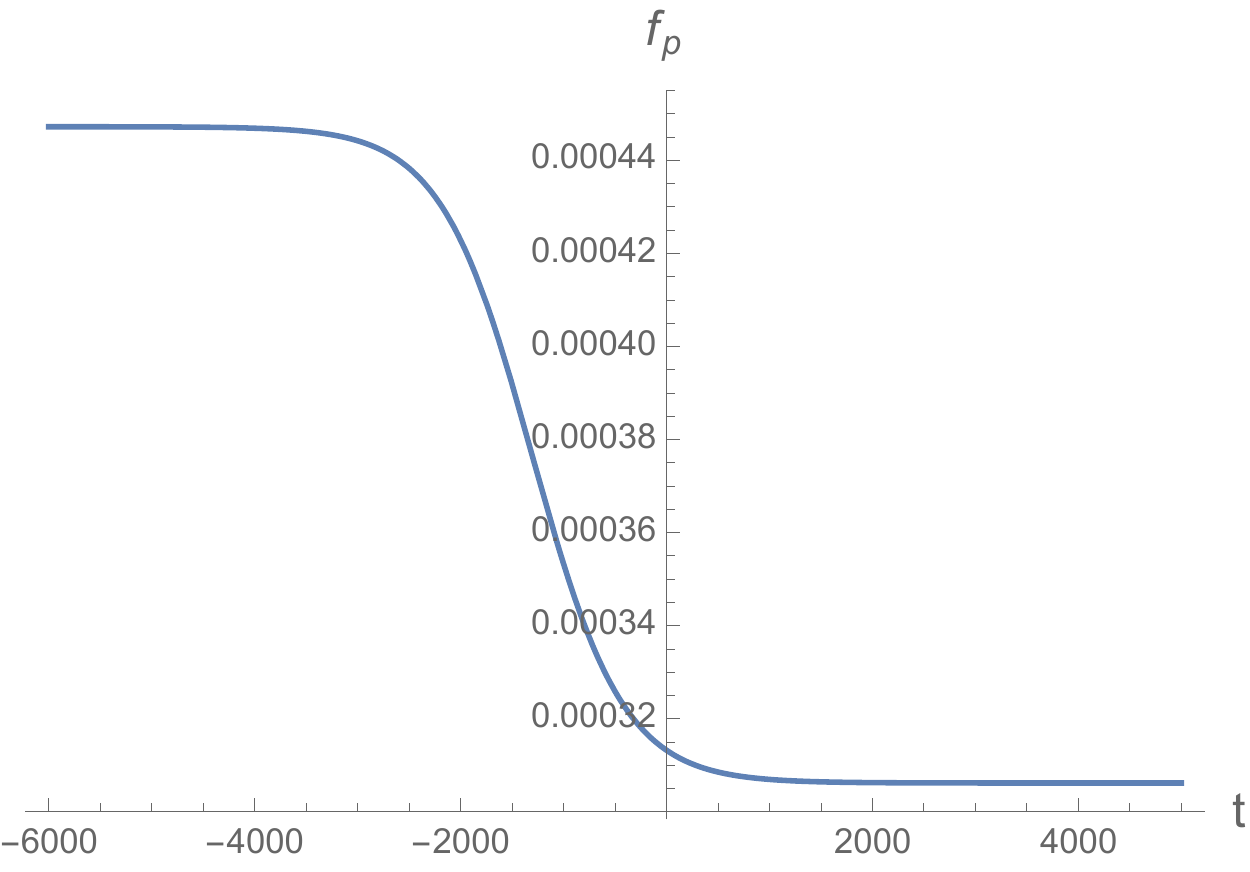}}
\end{subfigure}
\begin{subfigure}{.5\textwidth}
  \centering
  \scalebox{1}{\includegraphics[width=.8\linewidth]{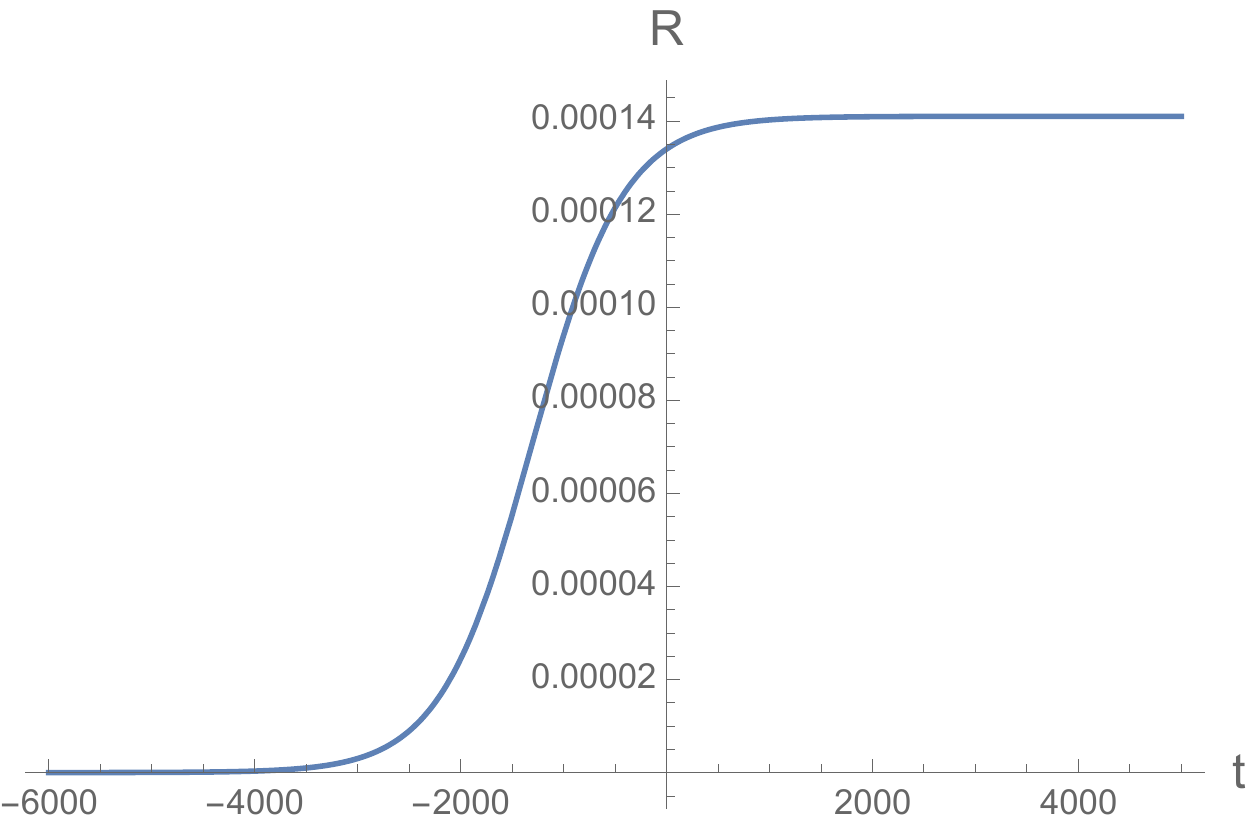}}
\end{subfigure}
\caption{RG flow  for $\lambda=0.001,\ \tilde\epsilon=0.1$: In the UV limit ($t \rightarrow\infty$) the couplings flow to a point on the fixed circle. In the IR limit ($t \rightarrow-\infty$) they flow to a fixed point where the two sectors are decoupled.}
\label{RG flows of fp-R}
\end{figure}

With the above comments in mind, we choose the marginally relevant deformation to be
\beq
\delta f_p=-\delta R= \frac{R_0}{2}\tilde\epsilon
\eeq
at a reference energy scale $\Lambda$. We take $0<\tilde\epsilon< 1$. The RG flow generated by the 1-loop beta functions leads to the following values of the couplings at a scale $\mu$: 
\begin{equation}
\begin{split}
&f_p=f_{0p}+\frac{R_0}{2}\Bigg[\frac{(1-k)\Big(1-\tanh(8R_0 t)\Big)}{1+k\tanh(8R_0 t)}\Bigg]\ ,\ R=\frac{R_0}{2}\Bigg[\frac{(1+k)\Big(1+\tanh(8R_0 t)\Big)}{1+k\tanh(8R_0 t)}\Bigg]\ ,\\
\end{split}
\end{equation}
where $t\equiv\ln(\mu/\Lambda)$ and $k\equiv 1-\tilde \epsilon$. In figure \ref{RG flows of fp-R}, a graphical plot of this RG flow is given for $\lambda=0.001,\ \tilde\epsilon=0.1$.
Note that the theory remains weakly coupled throughout the RG flow. In the deep UV, it flows to a  point on the   fixed circle as expected. In the deep IR, it flows to a fixed point where the two sectors are decoupled and $f_1=f_2={2\sqrt{6} +  \sqrt{18\sqrt{6}-39}\ov 16}\lam$. Due to the rotational symmetry in the $f_m$-$\zeta$ plane, this behavior of the RG flow is identical for marginally relevant deformations of all points on the fixed circle. In particular, when $r<r_\text{max}$, it holds for the deformations of the fixed points demonstrating thermal order in the second sector. The RG flows triggered by these deformations lead  to the IR fixed point where the two sectors are decoupled. As we have mentioned earlier, the baryon symmetries in both the sectors are unbroken at any nonzero temperature for this IR fixed point. Therefore, for the systems that flow from the UV fixed points with thermal order to this IR fixed point without thermal order, there must be a  transition from a disordered phase to an ordered phase as the temperature is increased. We will now determine the critical temperatures at which this phase transition takes place in these systems. 

\subsection{Estimates of the critical temperatures}
\label{subsec: critical temp estimate}

To determine the critical temperatures, we will  consider the thermal effective potential of the scalar fields. At leading order in perturbation theory,  this potential has both quadratic and quartic terms. The quartic terms are as follows:
\beq
V_{\text{quartic}}=16\pi^2\Bigg[\sum_{i=1}^2 \frac{h_i}{N_{ci}}\text{Tr}[(\Phi_i^T\Phi_i)^2]+\sum_{i=1}^2 \frac{f_i}{N_{ci}^2}\Big(\text{Tr}[\Phi_i^T\Phi_i]\Big)^2+\frac{2\zeta}{N_{c1} N_{c2}}\text{Tr}[\Phi_1^T\Phi_1]\text{Tr}[\Phi_2^T\Phi_2]\Bigg].
\label{quartic terms in effective potential}
\eeq
In appendix \ref{app: positivity of quartic terms} we have shown that these quartic terms are positive-definite throughout the RG flows that we have discussed in the previous subsection.  

The quadratic terms in the thermal effective potential are generated due to integration over the nonzero Matsubara modes of the different fields in the theory. These terms have the following structure
\beq
V_{\text{quadratic}}=\frac{1}{2}\sum_{i=1}^2 m_{\text{th},i}^2\text{Tr}[\Phi_i^T\Phi_i]
\eeq
where  $m_{\text{th},1}^2$ and $m_{\text{th},2}^2$ are the thermal masses (squared) of the scalar fields $\Phi_1$ and $\Phi_2$ respectively.  The   contributions of  1-loop  diagrams  to such thermal  masses (squared) were computed for a general 4-dimensional gauge theory in \cite{Weinberg:1974hy}. Using these general results, the 1-loop expressions of $m_{\text{th},1}^2$ and $m_{\text{th},2}^2$ were derived in \cite{Chaudhuri:2020xxb}. These expressions reduce to the following forms at the planar limit:
\begin{equation}
\begin{split}
&m_{\text{th},1}^2=\frac{16\pi^2 T^{2}}{3}\Bigg[2h_1+f_1+r\zeta+\frac{3}{8} \lambda_1\Bigg],\\
&m_{\text{th},2}^2=\frac{16\pi^2 T^{2}}{3}\Bigg[2 h_2+f_2+\frac{\zeta}{r}+\frac{3}{8} \lambda_2\Bigg],\\
\end{split}
\label{thermal mass expressions}
\end{equation}
where $T$ is the temperature.  These thermal masses (squared) quantify the behavior of the effective potential of the scalar fields in the neighborhood of the point $\Phi_1=\Phi_2=0$, viz., the origin of the field space. Their signs determine the fates of the baryon symmetries at any given  temperature. If either $m_{\text{th},1}^2$ or $m_{\text{th},2}^2$ is negative then the minimum of the effective potential cannot be at the origin of the field space. In fact, the positivity of the tree-level quartic terms in the potential ensures the existence of a minimum of the potential away from the origin. This would imply the spontaneous breaking of at least one of the baryon symmetries. In particular, when $m_{\text{th},i}^2<0$, the baryon symmetry in the $i^{\text{th}}$ sector would be spontaneously broken \cite{Chaudhuri:2020xxb}. 
 
To track the effective potential at different temperatures we will set the renormalization scale  $\mu=T$. Let us  explain this choice of the renormalization scale.  
Our aim  is to check perturbatively whether the thermal effective potential of the scalar fields has  a minimum away from the origin. When both the thermal masses (squared) are positive, the minimum of the potential is at the origin. In this case, as long as we are sufficiently away from the critical temperature, we will have $m_{\text{th},1}^2, m_{\text{th},2}^2\sim \frac{16\pi^2}{3}\lambda T^2$.\footnote{This is due to the fact that the different couplings are at most $O(\lambda)$ throughout the RG flows discussed in the previous subsection. From the expressions given in \eqref{thermal mass expressions}, one can see that if $r$ is  too small or too large then the magnitude of one of the thermal masses (squared) can become much larger than $\frac{16\pi^2 }{3} \lambda T^2$. In this paper, we will work in a regime where this is not the case.} At higher orders in the perturbative expansion of the potential near the origin, there are terms which  contain the logarithm of the ratio of  $m_{\text{th},i}^2$ and   $\mu^2$. The lowest order term of this kind is the Coleman-Weinberg term \cite{Coleman:1973jx, Lee:1974fj, Gould:2021oba}.  By setting $\mu=T$ we ensure that such logarithmic terms are suppressed by at least a factor of $\lambda\ln \lambda\ll 1$ compared to the leading order terms that we are retaining.\footnote{This suppression persists even when $T$ is very small.} A similar argument can be given when either of the thermal masses (squared) is negative and the minimum of the potential is away from the origin. In this case, again as long as we are sufficiently away from the critical point, we will have $|m_{\text{th},i}^2|\sim \frac{16\pi^2}{3}\lambda T^2$ irrespective of the sign of $m_{\text{th},i}^2$.\footnote{In fact, this behavior of the thermal masses (squared) would persist in the high temperature limit.} Comparing this behavior of the thermal masses (squared) with the quartic terms given in \eqref{quartic terms in effective potential}, one can check that if there is symmetry-breaking in the $i^{\text{th}}$ sector, then the value of $\Phi_i$ (appropriately normalized by $N_{ci}$) at the minimum would be of the same order as the temperature scale $T$. To be more precise, if  $(\Phi_i)_0$ is the value of $\Phi_i$ at the minimum, then  $\frac{\sqrt{\text{Tr}[(\Phi_i)_0^T(\Phi_i)_0]}}{N_{ci}}\sim  T$.\footnote{See \cite{Chaudhuri:2020xxb} for the explicit forms of such thermal expectation values of the scalar fields in the UV CFTs. Similar expressions would hold in the models that we are are studying presently.}
We want the perturbative expansion of the potential to be valid at this scale. At higher orders in this expansion, there are terms which  contain the logarithm of the ratio of $\lambda$ times the square of this scale and $\mu^2$. 
By taking $\mu=T$, we again ensure that such logarithmic corrections are small.
So we can rely on the perturbative analysis sufficiently away from the critical temperature on both sides of this temperature. To come up with an estimate of the critical temperature, we will interpolate the thermal masses (squared) in the intermediate interval with the expressions given in \eqref{thermal mass expressions} and search for the point at which one of the thermal masses (squared) vanishes.

One may be able to improve the above-mentioned perturbative analysis by choosing  $\mu$ to be some $O(1)$ numerical coefficient times $T$.  Such a numerical factor can be  absorbed in the reference energy scale $\Lambda$ by rescaling it appropriately while doing thermal perturbation theory.  Due to the ambiguity in the value of this numerical factor, there would be  theoretical uncertainties in the critical temperatures of the phase transitions that we will study.\footnote{Such theoretical uncertainties were discussed in \cite{Croon:2020cgk} for first order phase transitions in some models.}  Henceforth, we will ignore this subtlety as our aim is to find the qualitative nature of the phases in different temperature regimes and to determine the critical temperatures for the phase transitions in terms of the unspecified reference  energy scale $\Lambda$.

Having provided the rationale for choosing $\mu=T$ in the effective potential,  let us now look at the  behavior of the thermal masses at different temperature scales. For this it is convenient to define the  dimensionless quantity 
\beq
\tilde m_{\text{th},i}^2\equiv \frac{3}{16\pi^2 T^{2}}m_{\text{th},i}^2.
\eeq
 Substituting the values of the the different couplings along the RG flow triggered by the marginally relevant deformation, we get 
\begin{equation}
\begin{split}
\tilde m_{\text{th},1}^2=&\frac{R_0}{2}(1+k)\Big(r\cos\theta+\sin\theta-1\Big)C(t)+\Big(\frac{3}{4}\lambda+R_0\Big),\\
\tilde m_{\text{th},2}^2=&\frac{R_0}{2}(1+k)\Big(\frac{\cos\theta}{r}-\sin\theta-1\Big)C(t)+\Big(\frac{3}{4}\lambda+R_0\Big),\\
\end{split}
\end{equation}
where $\theta$ is the angular location of the UV fixed point on the conformal manifold and $C(t)$ is the following function of $t\equiv\ln(\frac{T}{\Lambda})$:
\beq
C(t)\equiv \frac{1+\tanh(8 R_0 t)}{1+k\tanh(8R_0 t)}.
\eeq
Note that the parameter $k=1-\tilde\epsilon$ controls the behavior of the thermal masses at different temperatures. As $\tilde\epsilon\rightarrow 0$, $C(t)\rightarrow 1$ and the rescaled thermal masses (squared), $\tilde m_{\text{th},1}^2$ and $\tilde m_{\text{th},2}^2$, stop changing with the temperature. In this limit, they just reduce to the rescaled thermal masses (squared) of the  points  on the conformal manifold which were analyzed in \cite{Chaudhuri:2020xxb}. 

\begin{figure}[H]
\begin{subfigure}{.5\textwidth}
  \centering
  \scalebox{1}{\includegraphics[width=.8\linewidth]{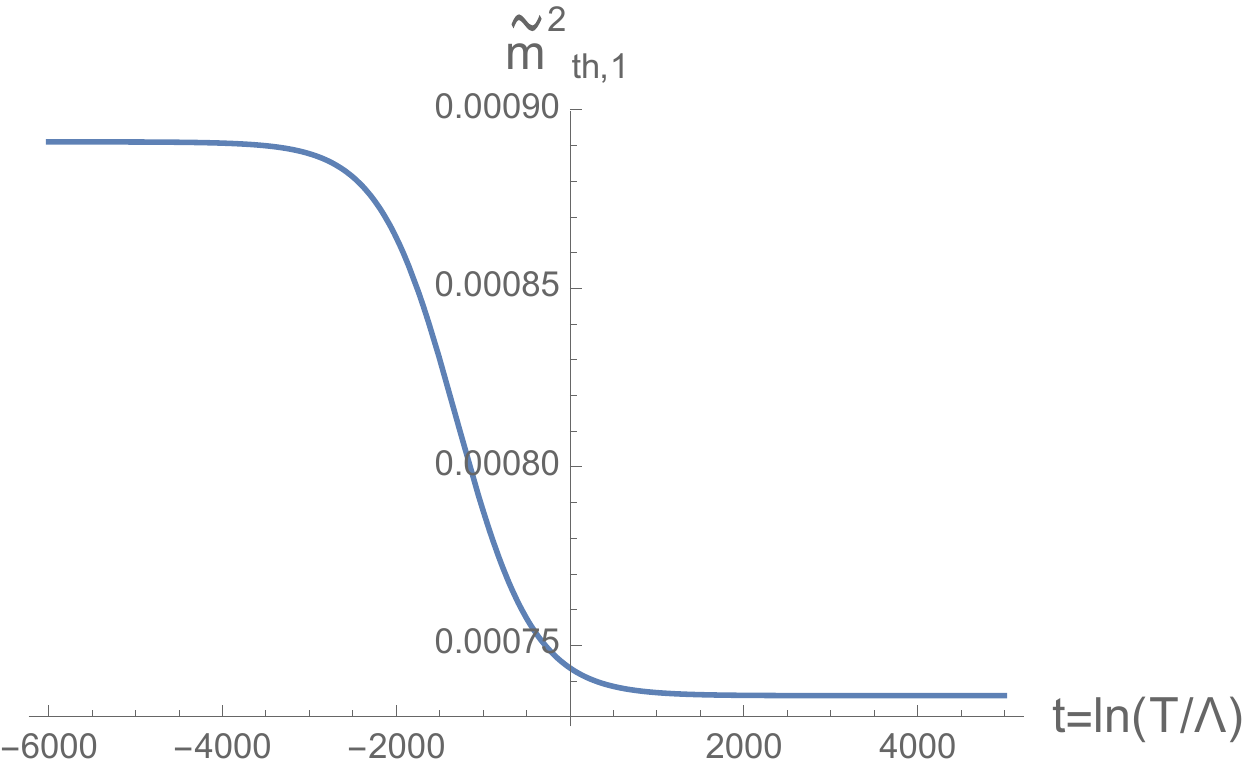}}
\end{subfigure}
\begin{subfigure}{.5\textwidth}
  \centering
  \scalebox{1}{\includegraphics[width=.8\linewidth]{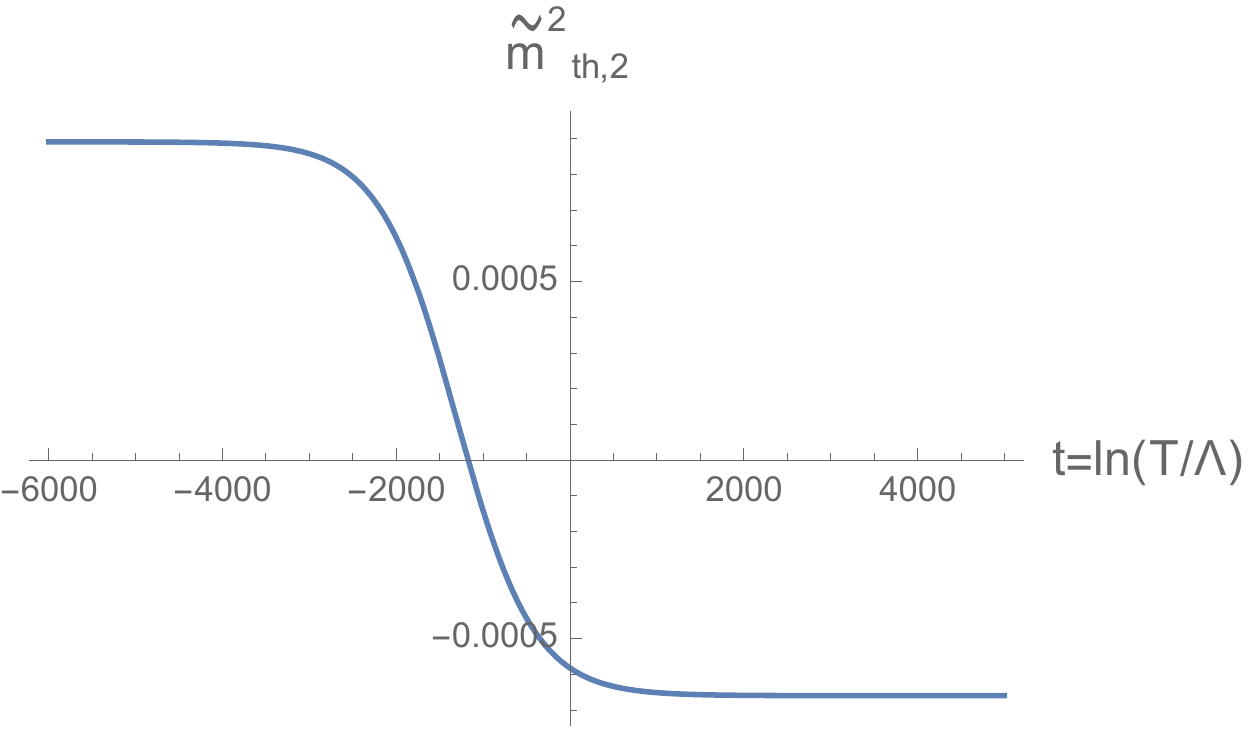}}
\end{subfigure}
\caption{Plots of the rescaled thermal masses  (squared) against  $t=\ln(T/\Lambda)$   for $r=0.1$, $\theta=\pi$, $\lambda=0.001$, $\tilde\epsilon=0.1$:  In the low temperature regime, both $\tilde m_{\text{th},1}^2$ and $\tilde m_{\text{th},2}^2$ saturate at positive values. This indicates that the  baryon symmetries in both the sectors are unbroken in this regime. In the high temperature limit, $\tilde m_{\text{th},1}^2$ saturates at a positive value, while $\tilde m_{\text{th},2}^2$ saturates at a negative value. This means that the baryon symmetry in the second sector remains persistently broken above a critical temperature. This critical temperature ($T_c$)  is given by $t_c\equiv\ln(T_c/\Lambda)\approx -1172.1$.}
\label{thermal masses plot}
\end{figure}

\begin{figure}
 \centering
\includegraphics[scale=0.6]{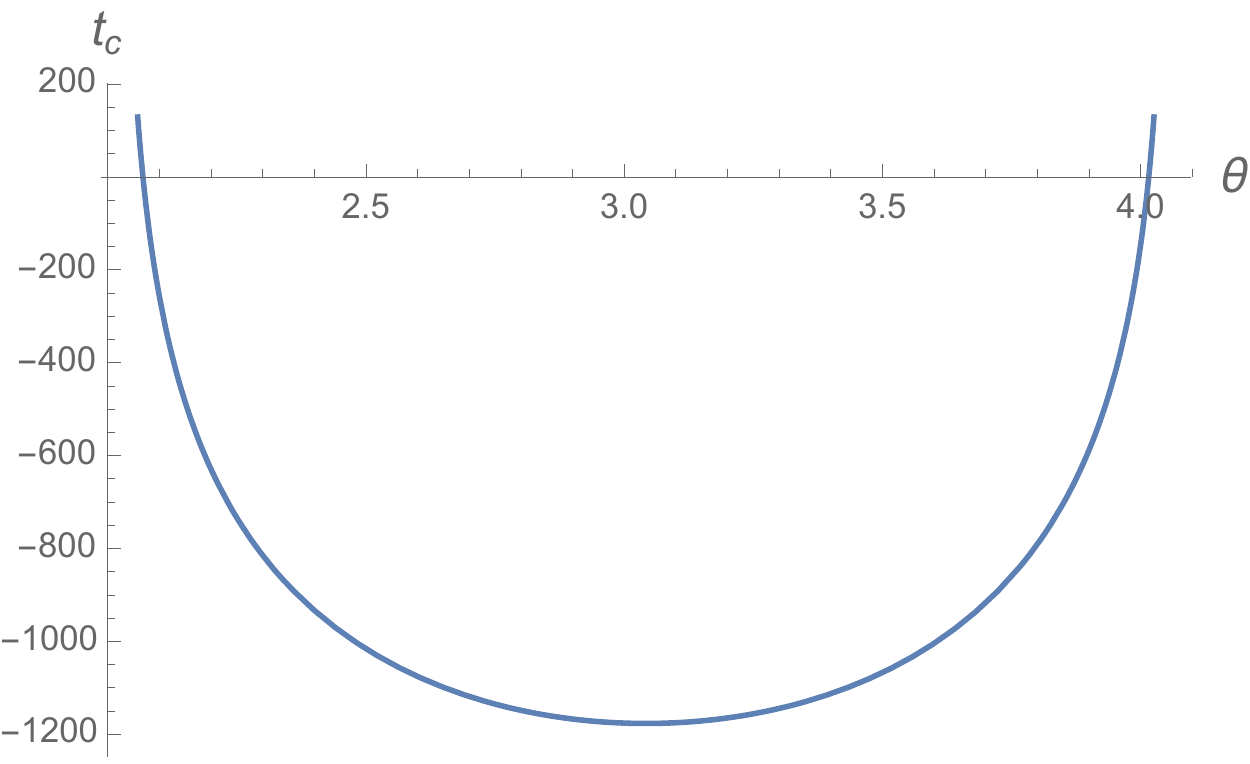}\,
\caption{Plot of $t_c$ against $\theta$ for $r=0.1,\lambda=0.001, \ \tilde\epsilon=0.1$:  The critical temperature  increases sharply near the edges of the interval $(\theta_1,\theta_2)$, i.e., $t_c\rightarrow\infty$ as $\theta\rightarrow\theta_1$ or $\theta\rightarrow\theta_2$.}
\label{tc vs theta}
\end{figure}

To study the variations of the thermal masses (squared) with the temperature concretely, we choose the UV fixed point to be the point where $\theta=\pi$, or equivalently, $f_m=0,\ \zeta=-R_0$. As can be verified from \eqref{domain of fixed points with thermal order}, this fixed point has thermal order in the second sector only when $r< r_0\approx 0.188$. So to stay in this regime, we choose $r=0.1$. To keep the values of the couplings and the deformations small we choose  $\lambda=0.001$ and $\tilde \epsilon=0.1$ as before. For this set of parameters, the plots of the rescaled thermal masses (squared) at different temperature scales are given in figure \ref{thermal masses plot}.
As one can see form these plots, both  $\tilde m_{\text{th},1}^2$ and $\tilde m_{\text{th},2}^2$ decrease monotonically with increase in the temperature. However,  $\tilde m_{\text{th},1}^2$ remains positive at all temperatures indicating the absence of thermal order in the first sector. On the other hand, $\tilde m_{\text{th},2}^2$ starts off from a positive value at low temperatures, but eventually becomes negative at high temperatures indicating a  transition to an ordered phase. The critical temperature at which this transition happens is where $\tilde m_{\text{th},2}^2=0$. This behavior is similar for all the systems where the UV fixed point exhibits thermal order in the second sector. One can obtain a general expression of the critical temperature ($T_c\equiv \Lambda e^{t_c}$) for all these systems by solving the equation $\tilde m_{\text{th},2}^2=0$. We provide this expression below:
\beq
t_c=\frac{1}{8R_0}\tanh^{-1}(\rho),
\eeq
where
\beq
\rho\equiv\frac{-(1+k)\frac{R_0}{2r}\Big(-\cos\theta+r(1+\sin\theta)\Big)+(\frac{3\lambda}{4}+R_0)}{(1+k)\frac{R_0}{2r}\Big(-\cos\theta+r(1+\sin\theta)\Big)-k(\frac{3\lambda}{4}+R_0)}.
\eeq
From the above expression, we can  see that $t_c$ has a real value only when $\rho \in(-1,1)$. This puts a restriction on the UV fixed points for which there can be a phase transition. The allowed fixed points are precisely the ones where $\theta\in(\theta_1,\theta_2)$ with  $\theta_1$ and $\theta_2$ being the values given in \eqref{domain of fixed points with thermal order}. To study the variation of $t_c$ in this domain, we choose $r=0.1$, $\lambda=0.001$, $\tilde \epsilon=0.1$ and plot the values of $t_c$ for different values of $\theta$ in figure \ref{tc vs theta}.
As one can see, $t_c$ increases rapidly as one approaches the edges of the interval $(\theta_1,\theta_2)$. As $\theta\rightarrow\theta_1$ or $\theta\rightarrow\theta_2$, $t_c\rightarrow\infty$ thereby indicating the absence of the phase transition at the end points. For each point in the interval $\theta\in(\theta_1,\theta_2)$, there is a finite temperature at which the  system goes to an ordered phase characterized by the spontaneous breaking of the baryon symmetry and Higgsing of half  of the gauge bosons in the second sector. In the Veneziano limit, the system remains in this phase at all higher temperatures. 

\begin{figure}[H]
 \centering
  \scalebox{1.3}{\includegraphics[scale=0.6]{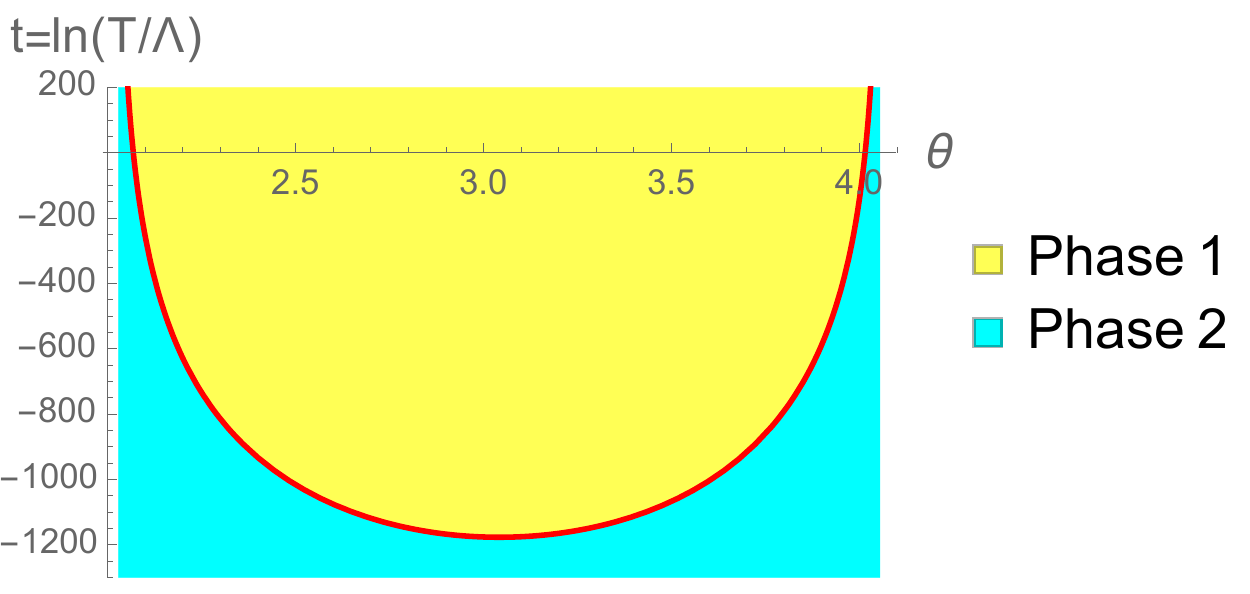}}\,
\caption{Phase diagram of the systems with $r=0.1$, $\lambda=0.001$, $\tilde \epsilon=0.1$: In Phase 1,  the baryon symmetry in the first sector is unbroken while the baryon symmetry in the second sector is spontaneously broken.  In Phase 2,  the baryon symmetries in both the  sectors are unbroken. The red line in this diagram is the line of critical points separating the two phases. It is the same as the curve given in figure \ref{tc vs theta}.}
\label{phase diagram massless case}
\end{figure}

The line of critical points shown in figure \ref{tc vs theta} separates the following two phases:
\begin{itemize}
\item Phase 1: The high temperature phase where the baryon symmetry in the first sector remains unbroken while the baryon symmetry in the second sector is spontaneously broken. 
\item Phase 2: The low temperature phase where the baryon symmetries in both the  sectors are unbroken.
\end{itemize}
In figure \ref{phase diagram massless case}, we show the phase diagram of the systems with $r=0.1$, $\lambda=0.001$, $\tilde \epsilon=0.1$. The phase diagrams for other values of $r$ in the domain $r<r_{\text{max}}$ would be similar to that shown in this figure.

Let us end this section with a remark on the zero temperature phase of the systems that we have been considering. At $T=0$, the thermal masses (squared) vanish and it is no longer possible to set the renormalization scale $\mu$ to $T$. In this case, one may wonder whether the Coleman-Weinberg terms can lead to a symmetry-broken phase. Such a symmetry-broken phase was indeed found in \cite{Coleman:1973jx} for massless scalar QED at zero temperature. As shown in \cite{Coleman:1973jx}, the quartic coupling in that  model is of the same order as the fourth power of the gauge coupling at certain renormalization scales. This leads to the Coleman-Weinberg term  becoming comparable to the quartic term slightly away from the origin which, in turn, allows for a perturbatively reliable minimum in this domain. In our case, however, the quartic couplings remain $O(\lambda)$ throughout the RG flow with $\lambda$ being related to the gauge couplings $g_i$ by $\lambda=\frac{N_{ci} g_i^2}{16\pi^2}$. The suppression of the Coleman-Weinberg terms compared to the quartic terms in this case is analogous to the same in massless $\phi^4$ theory where one does not get a perturbatively reliable minimum away from the origin at zero temperature \cite{Coleman:1973jx}. Therefore, in the systems that we are considering, we  expect the baryon symmetries in both the sectors to remain unbroken at zero temperature. We would like to establish this more rigorously in the future.

\section{Adding masses to the scalar fields}

\label{sec: Massive theory}

In the previous section we studied marginally relevant deformations of the  points on the fixed circle in the RDB model obtained by varying the double trace quartic couplings between the scalar fields. In this section we will  consider relevant deformations of the same fixed points obtained  by adding masses to the scalar fields.\footnote{The planar equivalence between the RDB model and the CDB model can be  extended to the case where the scalar fields are massive. These masses in the two dual theories are equal to each other.} For such deformations, the Lagrangian  has the following terms  in addition to  those given in \eqref{lagrangian}:
\beq
\mathcal{L}_{\text{mass}}=-\frac{1}{2}\sum_{i=1}^2 m_i^2\text{Tr}[\Phi_i^T\Phi_i],
\label{mass term}
\eeq
where $m_1^2$ and $m_2^2$ are the renormalized masses (squared) of the scalar fields in the two sectors. The 1-loop planar contributions to their RG flows (in the $\overline{\text{MS}}$ scheme)  are derived in appendix \ref{app: 1-loop beta function of masses}. These contributions are as follows:
\begin{equation}
\begin{split}
& \mu\frac{d m_1^2 }{d\mu}= \Big(16 h_1+8 f_1-3  \lambda_1\Big)m_1^2+8 r \zeta m_2^2,\\
& \mu\frac{d m_2^2 }{d\mu}= \Big(16 h_2+8 f_2-3  \lambda_2\Big)m_2^2+\frac{8 \zeta}{r} m_1^2.
\end{split}
\end{equation}
Since we are working in a mass-independent renormalization scheme for the RG flows of the quartic couplings, we can freeze these couplings at their fixed point values and study the flows of the masses. Substituting the couplings by their  leading order values on the fixed circle as given in \eqref{conformal manifold:1-loop fixed point}  and switching to the polar coordinates in the $f_m$-$\zeta$ plane (see \eqref{polar coordinates}), we get
\begin{equation}
\begin{split}
& \mu\frac{d m_1^2 }{d\mu}= 8R_0 \Big(\sin\theta \ m_1^2+ r \cos\theta \ m_2^2\Big),\\
&  \mu\frac{d m_2^2 }{d\mu}=8R_0\Big(\frac{\cos\theta}{r} \ m_1^2- \sin\theta \ m_2^2\Big),
\end{split}
\end{equation}
where $R_0$ is the radius of the fixed circle whose value at leading order in $\lambda$  is given by\footnote{We remind the reader that $\lambda_1=\lambda_2\equiv\lambda=\frac{21-4x_f}{13 x_f-27}$ on the fixed circle.}
\beq
R_0=\Big(\frac{\sqrt{18\sqrt{6}-39}}{16}\Big)\lambda.
\label{fixed circle radius}
\eeq 
Solving the above equations, we get
\begin{equation}
\begin{split}
&m_1^2=\frac{1}{2}(1-\sin\theta) M_1^2\ e^{-8R_0 t}+\frac{1}{2}(1+\sin\theta) M_2^2\ e^{8R_0 t},\\ 
&m_2^2=\frac{1}{2r}\Big(- M_1^2\ e^{-8R_0 t}+M_2^2\ e^{8R_0 t}\Big)\cos\theta,
\end{split}
\label{renormalized masses}
\end{equation}
where $t\equiv\ln(\mu/\Lambda)$ with $\Lambda$ being a reference energy scale. $M_1^2$ and $M_2^2$ are integration constants having the dimension of mass$^2$. When both of them are nonzero, they determine the asymptotic behavior of the renormalized masses (squared) in the IR and UV regimes respectively. 

At any nonzero temperature $T$ there are additional contributions to the effective masses (squared) of the scalar fields from the the thermal masses (squared) which were discussed in the previous section. For the massive theories that we are considering presently, the expressions of the 1-loop thermal masses (squared) given in \eqref{thermal mass expressions} are approximately valid when $T^2\gg |m_i^2|$. These expressions are given below:
\begin{equation}
\begin{split}
&m_{\text{th},1}^2=\frac{16\pi^2 T^{2}}{3}\Bigg[2h_1+f_1+r\zeta+\frac{3}{8} \lambda_1\Bigg],\\
&m_{\text{th},2}^2=\frac{16\pi^2 T^{2}}{3}\Bigg[2 h_2+f_2+\frac{\zeta}{r}+\frac{3}{8} \lambda_2\Bigg].\\
\end{split}
\label{thermal masses expressions}
\end{equation}
 For the points on the fixed circle, these expressions reduce to
\begin{equation}
\begin{split}
 m_{\text{th},1}^2=&\frac{16\pi^2 T^{2}}{3}\Bigg[R_0\Big(r\cos\theta+\sin\theta\Big)+\frac{3}{4}\lambda\Bigg],\\
 m_{\text{th},2}^2=&\frac{16\pi^2 T^{2}}{3}\Bigg[R_0\Big(\frac{\cos\theta}{r}-\sin\theta\Big)+\frac{3}{4}\lambda\Bigg].\\
\end{split}
\label{thermal masses on the fixed circle}
\end{equation}
The overall effective masses (squared) of the scalar fields at the temperature  $T$ are
\begin{equation}
\begin{split}
& m_{\text{eff},1}^2=(m_1^2+ m_{\text{th},1}^2)|_{\mu=T},\ m_{\text{eff},2}^2=(m_2^2+ m_{\text{th},2}^2)|_{\mu=T}.
\end{split}
\label{effective masses expressions}
\end{equation}
These effective masses (squared) determine the behavior of the effective potential of the scalar fields in the neighborhood of the point $\Phi_1=\Phi_2=0$. A  negative value of $m_{\text{eff},i}^2$ indicates that the baryon symmetry in the $i^{\text{th}}$ sector is spontaneously broken.

Here, as in the previous section, we have set $\mu=T$ to avoid large logarithms from appearing in the thermal perturbative expansion of the effective potential of the scalar fields. However, now one must be careful as the validity of this  choice of the renormalization scale rests on the magnitudes of effective masses (squared) being $O(\frac{16\pi^2}{3}\lambda T^2)$.\footnote{The validity of the perturbation theory also requires $|m_i^2|_{\mu=T}|\ll T^2.$ However, note that this is automatically ensured when $|m_{\text{eff},i}^2|\sim \frac{16\pi^2}{3}\lambda T^2$ because 
\beq
|m_i^2|_{\mu=T}|=|m_{\text{eff},i}^2-m_{\text{th},i}^2|<|m_{\text{eff},i}^2|+|m_{\text{th},i}^2|\sim\frac{32\pi^2}{3}\lambda T^2\ll T^2.
\eeq
Conversely, when the magnitudes of the effective masses (squared) become much larger than $O(\frac{16\pi^2}{3}\lambda T^2)$, the renormalized masses (squared) can become comparable to $T^2$. In that case, the expressions of the thermal masses (squared) given in \eqref{thermal masses on the fixed circle} are no longer reliable.}
At sufficiently low temperatures, the contributions of the renormalized masses (squared) will drive the effective masses (squared) out of this regime. So, unlike the previous section, here we cannot study the phase of the system down to arbitrarily low temperatures. However, we will show that in some cases, the renormalized masses can induce a phase transition at  scales where the perturbative analysis is reliable, i.e., near the critical temperatures corresponding to these phase transitions, the effective masses (squared) remain  $O(\frac{16\pi^2}{3}\lambda T^2)$. In the rest of this section, we will use the term `low temperatures' to mean temperatures slightly below such critical temperatures where the perturbative analysis is valid. We hope the reader will not be misled by this laxity.

The above-mentioned phase transitions  and the corresponding critical temperatures depend on the magnitudes and the signs of two mass$^2$ scales $M_1^2$ and $M_2^2$ which were introduced in \eqref{renormalized masses}. To study these phase transitions let us define the following two dimensionless quantities which are obtained via dividing $M_1^2$ and $M_2^2$ by  $\frac{32\pi^2}{3}\Lambda^2$:
 \beq
c_1\equiv \frac{3}{32\pi^2} \frac{M_1^2}{\Lambda^2} ,\  c_2\equiv \frac{3}{32\pi^2} \frac{M_2^2}{\Lambda^2}.
 \eeq
It is also convenient to switch to the following dimensionless quantities obtained via dividing  the effective masses (squared) by $\frac{16\pi^2}{3}T^2$ :
\begin{equation}
\tilde m_{\text{eff},1}^2\equiv\frac{3}{16\pi^2}\frac{m_{\text{eff},1}^2}{T^2},\ \tilde m_{\text{eff},2}^2\equiv\frac{3}{16\pi^2}\frac{m_{\text{eff},2}^2}{T^2}.
\end{equation}
In the regime where $m_i^2|_{\mu=T}\ll T^2$ and the expressions of the thermal masses (squared) given in \eqref{thermal masses on the fixed circle} are reliable, the rescaled effective masses (squared) defined above have the following values:

  \begin{equation}
\begin{split}
&\tilde m_{\text{eff},1}^2=(1-\sin\theta)  c_1\ e^{-(2+8R_0) t}+(1+\sin\theta) c_2\ e^{-(2-8R_0) t}+ R_0\Big(r\cos\theta+\sin\theta\Big)+\frac{3}{4}\lambda,\\ 
&\tilde m_{\text{eff},2}^2=\frac{1}{r}\Big(- c_1\ e^{-(2+8R_0) t}+c_2\ e^{-(2-8R_0) t}\Big)\cos\theta+R_0\Big(\frac{\cos\theta}{r}-\sin\theta\Big)+\frac{3}{4}\lambda.
\end{split}
\label{rescaled effective masses: general case}
\end{equation}
In the high temperature limit (i.e., large $t$), the rescaled effective masses (squared) are dominated by the contributions of the thermal masses (squared) and we have
  \begin{equation}
\begin{split}
&\tilde m_{\text{eff},1}^2 \xrightarrow{t\rightarrow\infty} R_0\Big(r\cos\theta+\sin\theta\Big)+\frac{3}{4}\lambda,\\ 
&\tilde m_{\text{eff},2}^2 \xrightarrow{t\rightarrow\infty}  R_0\Big(\frac{\cos\theta}{r}-\sin\theta\Big)+\frac{3}{4}\lambda.
\end{split}
\end{equation}
These high temperature behaviors of the effective masses (squared) are exactly as they would be for the UV CFTs. Therefore, for the systems  where $r< r_{\text{max}}=\sqrt{\frac{6\sqrt{6}-13}{61-6\sqrt{6}}}$ and $\theta\in(\theta_1,\theta_2)$ with $\theta_1$ and $\theta_2$ being the values given in \eqref{domain of fixed points with thermal order},  the baryon symmetry in the first sector remains unbroken in the high temperature regime, whereas the baryon symmetry in the second sector is persistently broken in the same regime. In the rest of this section  we will restrict our attention to these systems which exhibit thermal order in the second sector at the high temperature limit. As the temperature is lowered in these systems, i.e., as $t$ decreases, the contributions of the renormalized masses (squared) start becoming comparable to the thermal masses  (squared). This is what allows for phase transitions in the system.  From the expressions given in \eqref{rescaled effective masses: general case} one can  see that the signs of $c_1$ and $c_2$ (or equivalently, the signs of the mass$^2$ scales $M_1^2$ and $M_2^2$) play the key role in determining these phase transitions. We will study these phase transitions in the following 2 cases:
\begin{itemize}
\item Case 1: $M_1^2\neq0, M_2^2=0$,
\item Case 2:  $M_1^2=0, M_2^2\neq 0$. 
\end{itemize}
We will see that for each of these cases, there are distinct patterns of phase transitions depending on the signs of the nonzero $M_i^2$. We will discuss the low temperature phases for the different signs of the nonzero $M_i^2$ in each case as separate subcases. Later, in subsection \ref{subsec:Massive case summary} we will summarize these results and comment on the possible phase transitions when both $M_1^2$ and $M_2^2$ are nonzero. 

\subsection{Case 1: $M_1^2\neq0, M_2^2=0$}
In this case the expressions of the rescaled effective masses (squared) reduce to the following forms:
  \begin{equation}
\begin{split}
&\tilde m_{\text{eff},1}^2=(1-\sin\theta)  c_1\ e^{-(2+8R_0) t}+ R_0\Big(r\cos\theta+\sin\theta\Big)+\frac{3}{4}\lambda,\\ 
&\tilde m_{\text{eff},2}^2=-\frac{1}{r}\cos\theta \ c_1\ e^{-(2+8R_0) t}+R_0\Big(\frac{\cos\theta}{r}-\sin\theta\Big)+\frac{3}{4}\lambda.
\end{split}
\label{effective masses : Case 1}
\end{equation}
We remind the reader that we are restricting our attention to the systems which exhibit thermal order in the second sector at high temperatures, i.e., the systems where $r< r_{\text{max}}$ and $\theta\in(\theta_1,\theta_2)$. As we have already mentioned in section \ref{sec: dbm review}, $\cos\theta$ is always negative for such systems. Moreover, the quantity $(1-\sin\theta)$ is  positive for all values of $\theta$ in this domain. Therefore, from the expressions given in \eqref{effective masses : Case 1} we can see that  in the low temperature regime where the renormalized masses (squared) dominate over the thermal masses (squared),  both the effective masses (squared) have the same sign. This sign depends on whether $c_1$ (or equivalently, $M_1^2$) is positive or negative.

\subsubsection{Subcase 1: $M_1^2>0$}

When $M_1^2>0$, both the effective masses (squared) are positive in the low temperature regime. This indicates that the baryon symmetries in both the sectors are unbroken in this regime. Therefore, there should be a phase transition at some critical temperature where the baryon symmetry in the second sector is spontaneously broken.

\begin{figure}[H]
  \centering
  \scalebox{0.5}{\includegraphics[width=.8\linewidth]{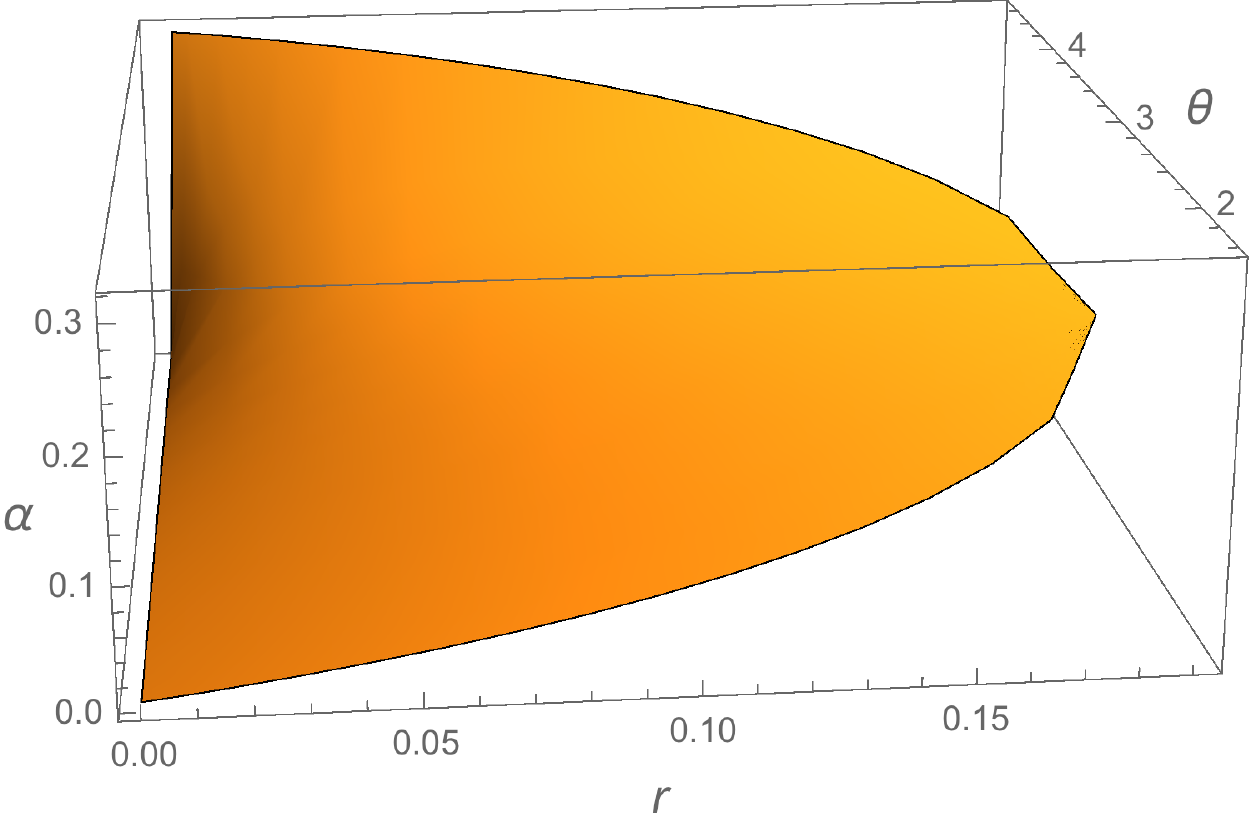}}
\caption{Plot of $\alpha= -\frac{r(1-\sin\theta)}{\cos\theta}$ against $r$ and $\theta$ in the domain $r<r_{\text{max}}$ and $\theta\in(\theta_1,\theta_2)$.}
\label{alpha plot}
\end{figure}

At the critical temperature ($T_c$) corresponding to this phase transition $m_{\text{eff},2}^2$  vanishes, i.e., the renormalized mass (squared) $m_2^2|_{\mu=T_c}$ exactly cancels the thermal mass (squared) $m_{\text{th},2}^2$ . Therefore,  we have 
\beq
|m_2^2|_{\mu=T_c}|\sim \frac{16\pi^2}{3}\lambda T_c^2\ll T_c^2.
\label{Case 1: Subcase 1: renormalized mass condition 1}
\eeq
Moreover, note that the two renormalized masses (squared) are related as follows
\beq
m_1^2|_{\mu=T_c}=\alpha m_2^2|_{\mu=T_c}
\label{Case 1: Subcase 1: relation between renormalized masses}
\eeq
where $\alpha\equiv -\frac{r(1-\sin\theta)}{\cos\theta}$. Since we are considering systems where $r\in(0,r_{\text{max}})$ and $\theta\in(\theta_1,\theta_2)$ with $\theta_1$  and $\theta_2$ satisfying $\frac{\pi}{2}<\theta_1<\theta_2<\frac{3\pi}{2}$, both the numerator and denominator in $\alpha$ are nonzero. In fact, the denominator can become small only when $\theta $ is close to $ \pi/2$ or $3\pi/2$. Such values of $\theta$ can lie in the range $(\theta_1,\theta_2)$ only when $r$ is close to zero. Therefore, in this case the factor of $r$ in the numerator of the expression of $\alpha$ ensures that $\alpha$ remains small despite the denominator having a small value. In figure \ref{alpha plot} we show the values of $\alpha$ for all values $(r,\theta)$ satisfying $r<r_{\text{max}}$ and $\theta_1<\theta<\theta_2$. We can see that $\alpha<1$ at all points in this domain.
 Thus, from \eqref{Case 1: Subcase 1: renormalized mass condition 1} and \eqref{Case 1: Subcase 1: relation between renormalized masses} we can conclude that 
\beq
|m_1^2|_{\mu=T_c}|\lesssim \frac{16\pi^2}{3}\lambda T_c^2\ll T_c^2
\eeq
for all these systems. Note that $|m_{1,2}^2|_{\mu=T_c}|\ll T_c^2$ means that the expressions of the thermal masses (squared) given in \eqref{thermal masses on the fixed circle} are reliable at the critical temperature. Moreover, the fact that $|m_1^2|_{\mu=T_c}|\lesssim \frac{16\pi^2}{3}\lambda T_c^2$ ensures that at the critical temperature $|m_{\text{eff},1}^2|\sim \frac{16\pi^2}{3} \lambda T_c^2$. This means that the perturbation theory with $\mu=T$ can be trusted at  temperatures near $T_c$. Hence, one can study the phase transition using this renormalization scale.

To study this phase transition with a concrete example, we plot the rescaled effective masses (squared) against $t\equiv \ln(T/\Lambda)$ for the case  where $r=0.1,\ \theta=\pi$, $\lambda=0.001$, $c_1=0.1$ and $c_2=0$ in figure \ref{effective masses: case 1, subcase 1}.
One can check that the choice of $r$ and $\theta$ is such that the UV fixed point exhibits thermal order in the second sector. 
\begin{figure}[H]
\begin{subfigure}{.5\textwidth}
  \centering
  \scalebox{1}{\includegraphics[width=.8\linewidth]{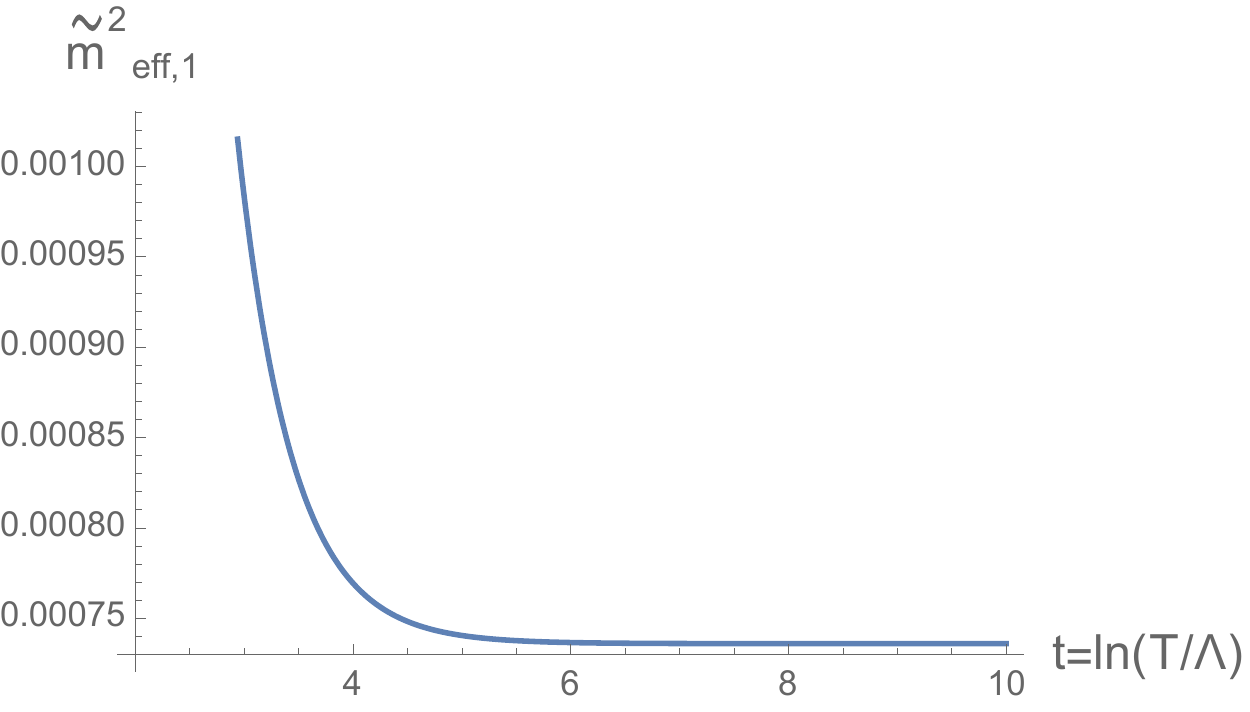}}
\end{subfigure}
\begin{subfigure}{.5\textwidth}
  \centering
  \scalebox{1}{\includegraphics[width=.8\linewidth]{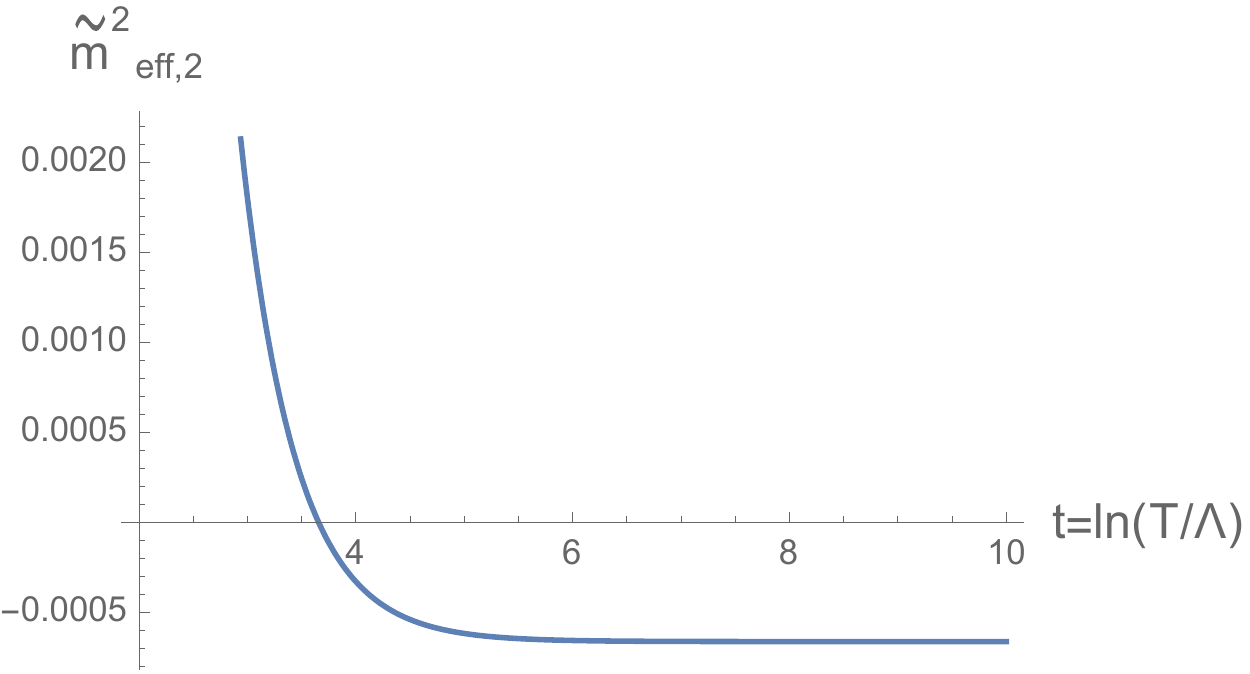}}
\end{subfigure}
\caption{Plots of the  rescaled effective masses (squared) against $t=\ln(T/\Lambda)$  for $r=0.1$, $\theta=\pi$, 
$\lambda=0.001$, $c_1=0.1$, $c_2=0$: Both $\tilde m_{\text{eff},1}^2$ and $\tilde m_{\text{eff},2}^2$ grow as the temperature decreases. In the high temperature limit, $\tilde m_{\text{eff},1}^2$ saturates at a positive value while $\tilde m_{\text{eff},2}^2$ saturates at a negative value. This indicates a phase transition at an intermediate temperature where the baryon symmetry in the second sector is broken. The critical temperature ($T_c$) corresponding to this phase transition is given by $t_c\equiv\ln(T_c/\Lambda)\approx 3.659$.}
\label{effective masses: case 1, subcase 1}
\end{figure}
Note that $\tilde m_{\text{eff},1}^2$ remains positive for all values of $t$. It grows as one goes to lower temperatures. At high temperatures, it saturates at a positive value. This means that the baryon symmetry in the first sector remains unbroken at all temperatures where the perturbative analysis is reliable. On the other hand,  $\tilde m_{\text{eff},2}^2$ also grows as the temperature is lowered, but at high temperatures it saturates at a negative value. This means that the baryon symmetry in the second sector is spontaneously broken at a critical temperature and then it remains broken at all higher temperatures.

 In order to determine this critical temperature, we can solve for the value of $t$ at which $\tilde m_{\text{eff},2}^2$ vanishes. Setting $\tilde m_{\text{eff},2}^2=0$ at $t=t_c=\ln(T_c/\Lambda)$, we get
\begin{equation}
\begin{split}
&   t_c=  -\frac{1}{2+8R_0}\ln\Bigg[\frac{r}{c_1\cos\theta}\Bigg\{R_0\Big(\frac{\cos\theta}{r}-\sin\theta\Big)+\frac{3}{4}\lambda\Bigg\}\Bigg].
\end{split}
\end{equation}
In figure \ref{tc : case 1, subcase 1}, we provide the plot of $t_c$ corresponding to different values of $\theta\in(\theta_1,\theta_2)$ for $\ r=0.1,\ \lambda=0.001,\ c_1=0.1,\ c_2=0$. 
\begin{figure}[H]
  \centering
  \scalebox{0.6}{\includegraphics[width=.8\linewidth]{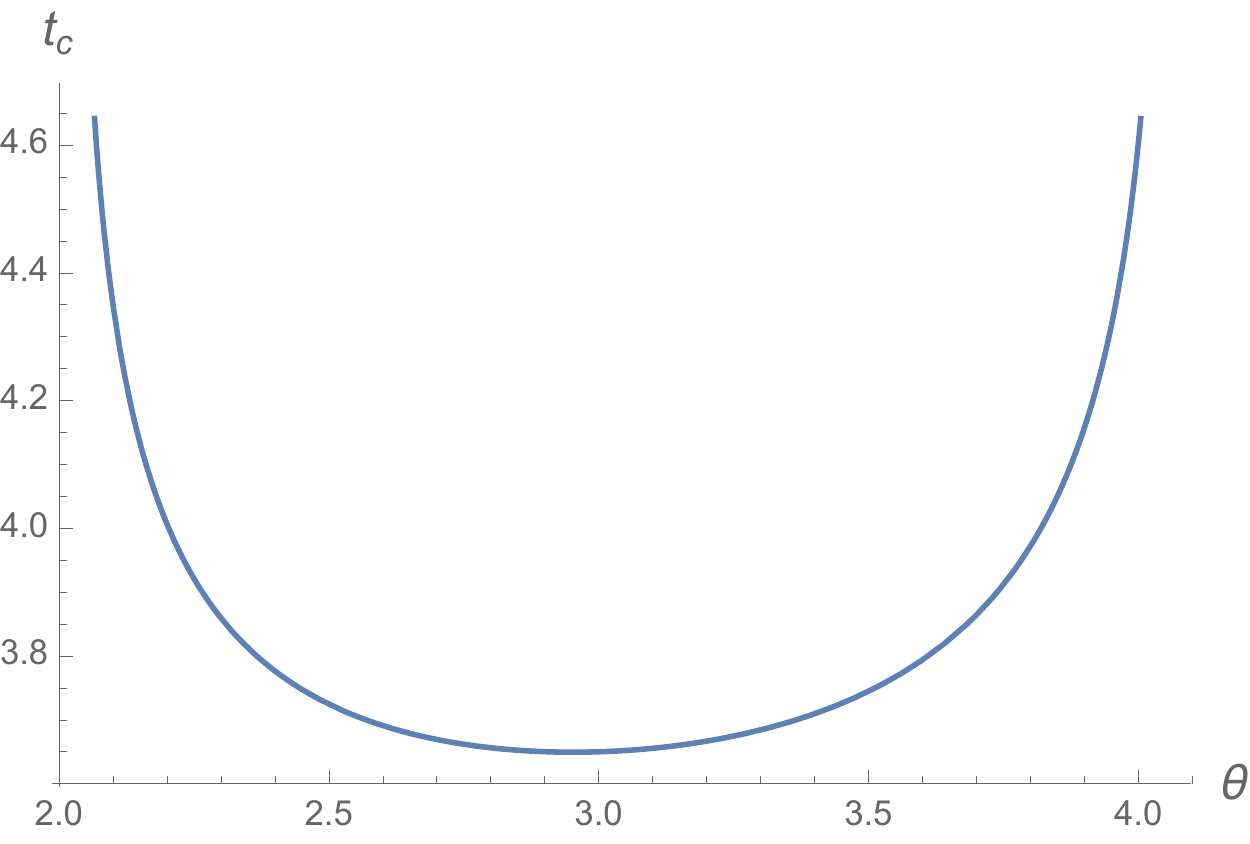}}
\caption{Plot of $t_c$  against $\theta$  for $r=0.1,\ \lambda=0.001, \ c_1=0.1,\ c_2=0$: The critical temperature  increases sharply near the edges of the interval $(\theta_1,\theta_2)$,i.e., $t_c\rightarrow\infty$ as $\theta\rightarrow\theta_{1,2}$.}
\label{tc : case 1, subcase 1}
\end{figure}
Just like what we saw for the critical temperatures corresponding to the phase transitions discussed in section \ref{sec: marginally relevant deformations:phase transitions}, the critical temperature in this case increases sharply as one approaches the edges of the domain $(\theta_1,\theta_2)$, and $t_c\rightarrow\infty$ as $\theta\rightarrow\theta_1$ or $\theta\rightarrow\theta_2$. This means that at these edges, the system no longer undergoes the aforementioned phase transition and the baryon symmetries in both the sectors remain unbroken at all temperatures where the perturbative analysis can be trusted.

\subsubsection{Subcase 2: $M_1^2<0$}

When $M_1^2<0$, both the effective masses (squared) are negative in the low temperature regime. This means that the baryon symmetries in both the sectors are spontaneously broken in this regime. On the other hand, in the high temperature regime, the baryon symmetry in the first sector is unbroken while the baryon symmetry in the second sector is spontaneously broken. Therefore, we expect a phase transition at some critical temperature where the baryon symmetry in the first sector is restored. 

At this critical temperature, the effective mass (squared) $m_{\text{eff},1}^2$ should vanish, i.e.,  $m_1^2|_{\mu=T_c}$ should cancel the contribution of the $m_{\text{th},1}^2$, and consequently 
\beq
|m_1^2|_{\mu=T_c}|\sim \frac{16\pi^2}{3} \lambda T_c^2\ll T_c^2.
\eeq 
Now,  we can again use the relation \eqref{Case 1: Subcase 1: relation between renormalized masses} between $m_1^2$ and $m_2^2$ to get
\beq
m_2^2|_{\mu=T_c}=\frac{m_1^2|_{\mu=T_c}}{\alpha}
\eeq
where $\alpha\equiv -\frac{r(1-\sin\theta)}{\cos\theta}$. As we had noted earlier, $\alpha<1$ for all the UV fixed points which we are considering, i.e., the fixed points where $r<r_{\text{max}}$ and $\theta\in(\theta_1,\theta_2)$. From the plot given in figure \ref{alpha plot} we can see that in certain regimes of $r$ and $\theta$, $\alpha$ can be much smaller than 1. In such regimes,  the magnitude of $m_2^2|_{\mu=T_c}$ will be much larger than $\frac{16\pi^2}{3}\lambda T_c^2$ making the perturbative analysis with $\mu=T$ unreliable near the critical temperature.

In order to reliably employ the perturbative analysis to study the phase transition, we  choose a system where $r=0.03$ and $\theta=4.52$. One can check that the  fixed point to which this system flows in the UV lies in the domain given in \eqref{domain of fixed points with thermal order}, i.e., it exhibits thermal order in the second sector. For this point, we have $\alpha\approx 0.311$, and hence
\beq
m_2^2|_{\mu= T_c}\approx 3.216\ m_1^2|_{\mu=T_c}.
\eeq
Therefore, near the critical temperature, we remain in a regime where $|m_{\text{eff},2}^2|\sim \frac{16\pi^2}{3}\lambda T^2$, and the perturbative analysis is reliable. 

In figure \ref{effective masses: case 1, subcase 2}, we plot  $\tilde m_{\text{eff},1}^2$ and $\tilde m_{\text{eff},2}^2$ against $t\equiv \ln(T/\Lambda)$ for the case where $r=0.03$, $\theta=4.52$, $\lambda=0.001$ $c1=-0.1$ and  $c_2=0$. 
\begin{figure}[H]
\begin{subfigure}{.5\textwidth}
  \centering
  \scalebox{1}{\includegraphics[width=.8\linewidth]{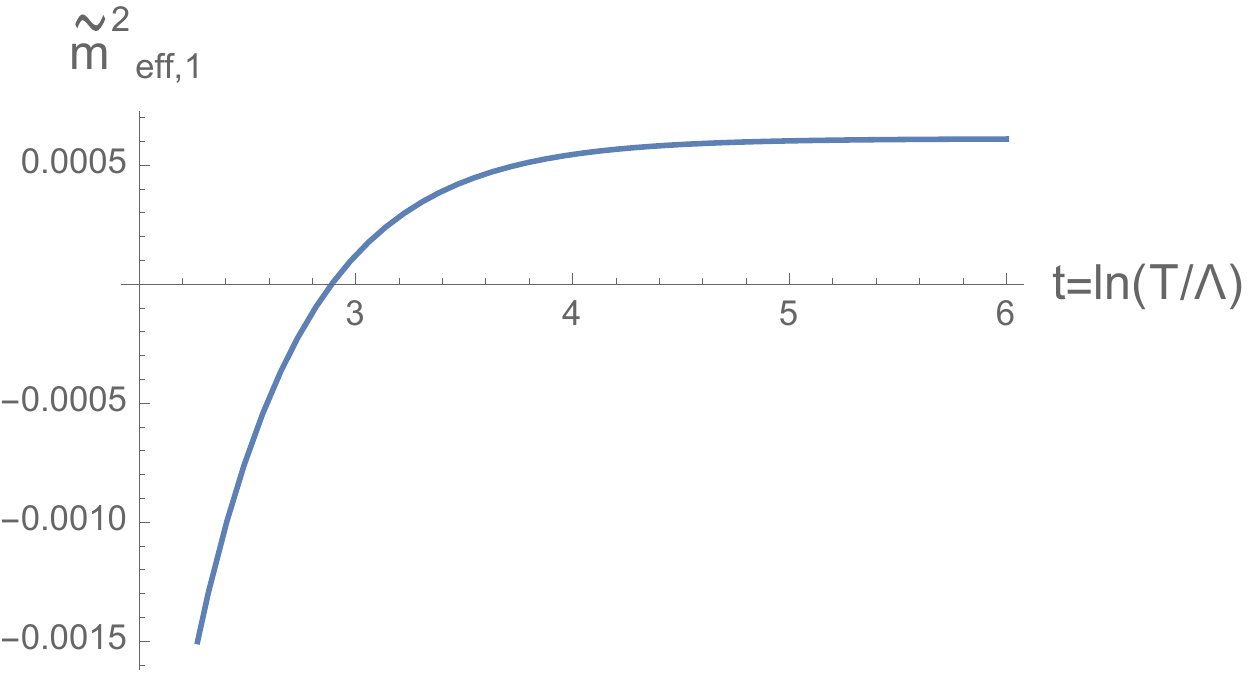}}
\end{subfigure}
\begin{subfigure}{.5\textwidth}
  \centering
  \scalebox{1}{\includegraphics[width=.8\linewidth]{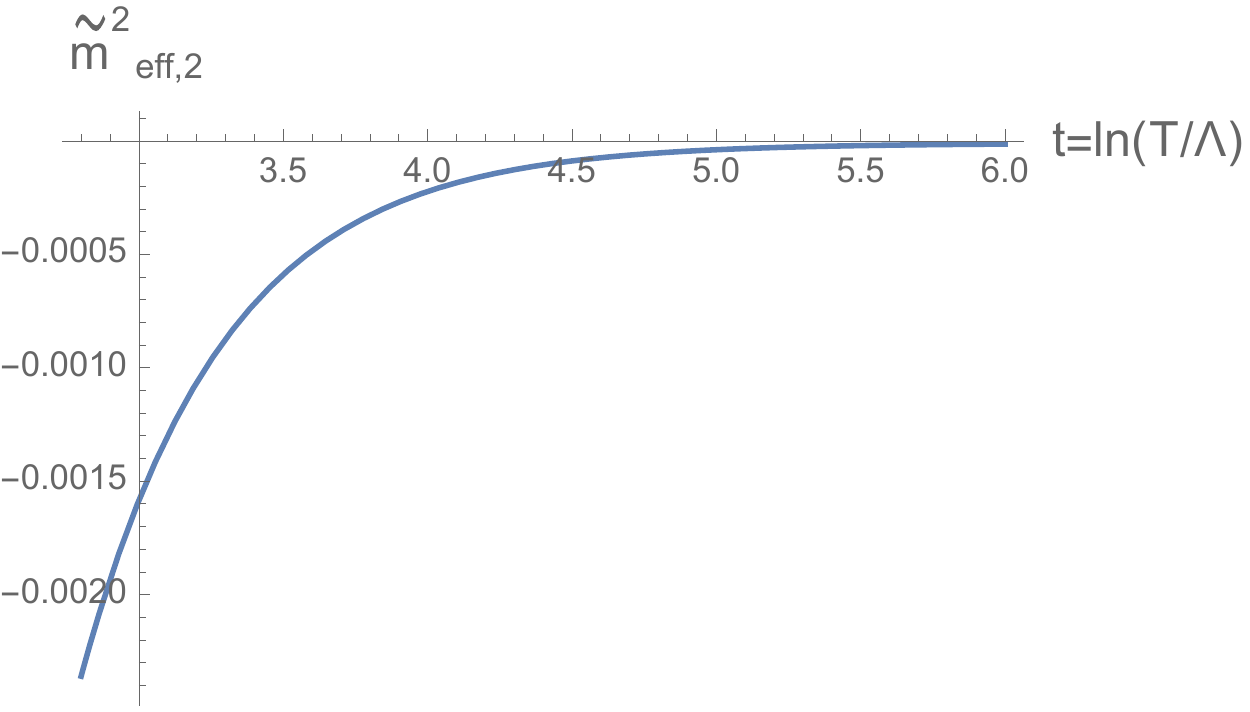}}
\end{subfigure}
\caption{Plots of the  rescaled effective masses (squared) against $t=\ln(T/\Lambda)$  for $r=0.03$, $\theta=4.52$, $\lambda=0.001$, $c_1=-0.1$, $c_2=0$:  Both $\tilde m_{\text{eff},1}^2$ and $\tilde m_{\text{eff},2}^2$ are negative in the low temperature regime, and their magnitudes grow as the temperature is decreased. In the high temperature limit, $\tilde m_{\text{eff},1}^2$ saturates at a positive value while $\tilde m_{\text{eff},2}^2$ saturates at a negative value. This indicates that the baryon symmetry in the second sector remains broken at all temperatures where the perturbative analysis can be trusted, while the baryon symmetry in the first sector is restored at a critical temperature. This critical temperature ($T_c$) is given by $t_c=\ln(T_c/\Lambda)\approx 2.889$.
}
\label{effective masses: case 1, subcase 2}
\end{figure}
Here $\tilde m_{\text{eff},2}^2$ remains negative for all values of $t$. Its magnitude grows as one goes to lower temperatures. At high temperatures, it saturates at a negative value. This means that the baryon symmetry in the second sector remains broken at all temperatures where the perturbative analysis is valid. The behavior of $\tilde m_{\text{eff},1}^2$ is quite similar to that of $\tilde m_{\text{eff},2}^2$ in the low temperature regime, i.e., in this regime $\tilde m_{\text{eff},1}^2<0$ and $|\tilde m_{\text{eff},1}^2|$  grows as the temperature is lowered. However, at high temperatures $\tilde m_{\text{eff},1}^2$ saturates at a positive value. This means that the baryon symmetry in the first sector is restored at a critical temperature. The value of of this critical temperature ($T_c$) is given by $t_c=\ln(T_c/\Lambda)\approx 2.889$.

\subsection{Case 2: $M_1^2=0, M_2^2\neq 0$}

In this case the expressions of the rescaled effective masses (squared) are
  \begin{equation}
\begin{split}
&\tilde m_{\text{eff},1}^2=(1+\sin\theta) c_2\ e^{-(2-8R_0) t}+ R_0\Big(r\cos\theta+\sin\theta\Big)+\frac{3}{4}\lambda,\\ 
&\tilde m_{\text{eff},2}^2=\frac{1}{r}\cos\theta\ c_2\ e^{-(2-8R_0) t}+R_0\Big(\frac{\cos\theta}{r}-\sin\theta\Big)+\frac{3}{4}\lambda.
\end{split}
\end{equation}
As we have  noted earlier,  $\cos\theta<0$ for the systems that we are considering, and $(1+\sin\theta)>0$ for all values of $\theta$ in this domain. Therefore, at sufficiently low temperatures, when the renormalized masses (squared) dominate over the thermal masses (squared), $\tilde m_{\text{eff},1}^2$ and $\tilde m_{\text{eff},2}^2$ have opposite signs. These signs depends on whether $c_2$ (or equivalently, $M_2^2$) is positive or negative. 

 \subsubsection{Subcase 1: $M_2^2>0$}
 
 When  $M_2^2>0$, i.e., $c_2>0$,  then $\tilde m_{\text{eff},1}^2>0$ and $\tilde m_{\text{eff},2}^2<0$ in the low temperature regime. This means that in this regime the baryon symmetry in  the first sector remains unbroken and the baryon symmetry in  the second sector is spontaneously broken. This is identical to the phase in the high temperature limit. Therefore, we expect the system to remain in the same phase at all temperatures in the domain of validity of the perturbative analysis. To verify this we plot $\tilde m_{\text{eff},1}^2$ and $\tilde m_{\text{eff},2}^2$ at different temperatures for $r=0.1$, $\theta=\pi$, $\lambda=0.001$, $c_1=0$ and $c_2=0.1$ in figure \ref{effective masses: case 2, subcase 1}. We can see that $\tilde m_{\text{eff},1}^2>0$ and  $\tilde m_{\text{eff},2}^2<0$  for all values of $t$. This indicates that the system remains in the same phase at all temperatures where perturbation theory can be trusted.
 \begin{figure}[H]
\begin{subfigure}{.5\textwidth}
  \centering
  \scalebox{1}{\includegraphics[width=.8\linewidth]{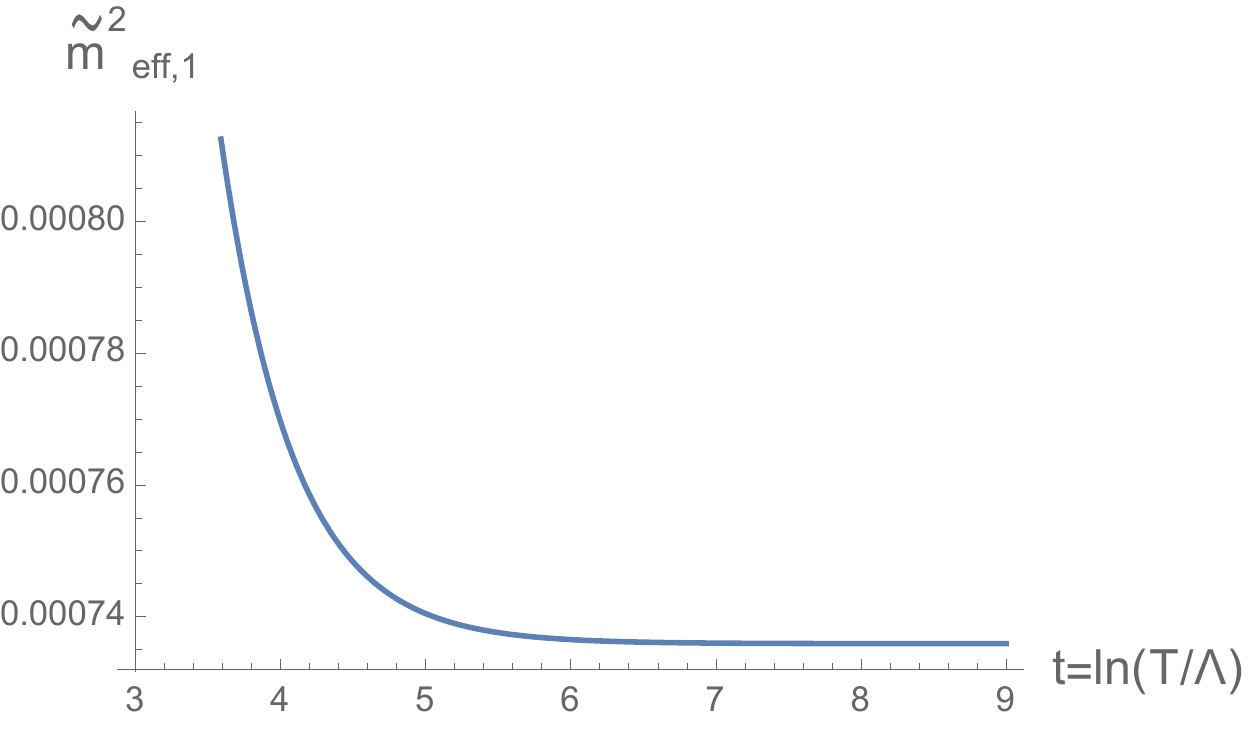}}
\end{subfigure}
\begin{subfigure}{.5\textwidth}
  \centering
  \scalebox{1}{\includegraphics[width=.8\linewidth]{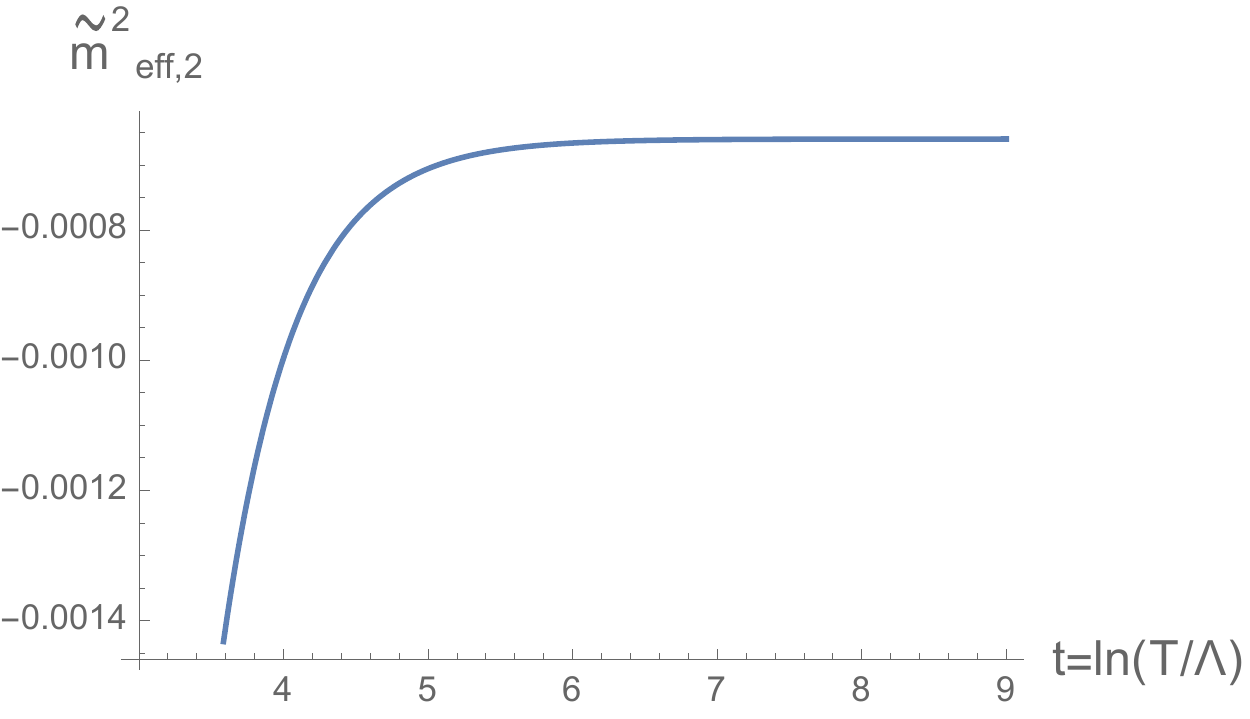}}
\end{subfigure}
\caption{Plots of the  rescaled  effective masses (squared) against $t=\ln(T/\Lambda)$  for $r=0.1$, $\theta=\pi$, $\lambda=0.001$, $c_1=0$, $c_2=0.1$:   $\tilde m_{\text{eff},1}^2>0$ and $\tilde m_{\text{eff},2}^2<0$ at all temperatures indicating the absence of any phase transition in the domain of validity of the perturbative analysis.
}
\label{effective masses: case 2, subcase 1}
\end{figure}

 \subsubsection{Subcase 2: $M_2^2<0$}
 
\begin{figure}[H]
  \centering
  \scalebox{0.5}{\includegraphics[width=.8\linewidth]{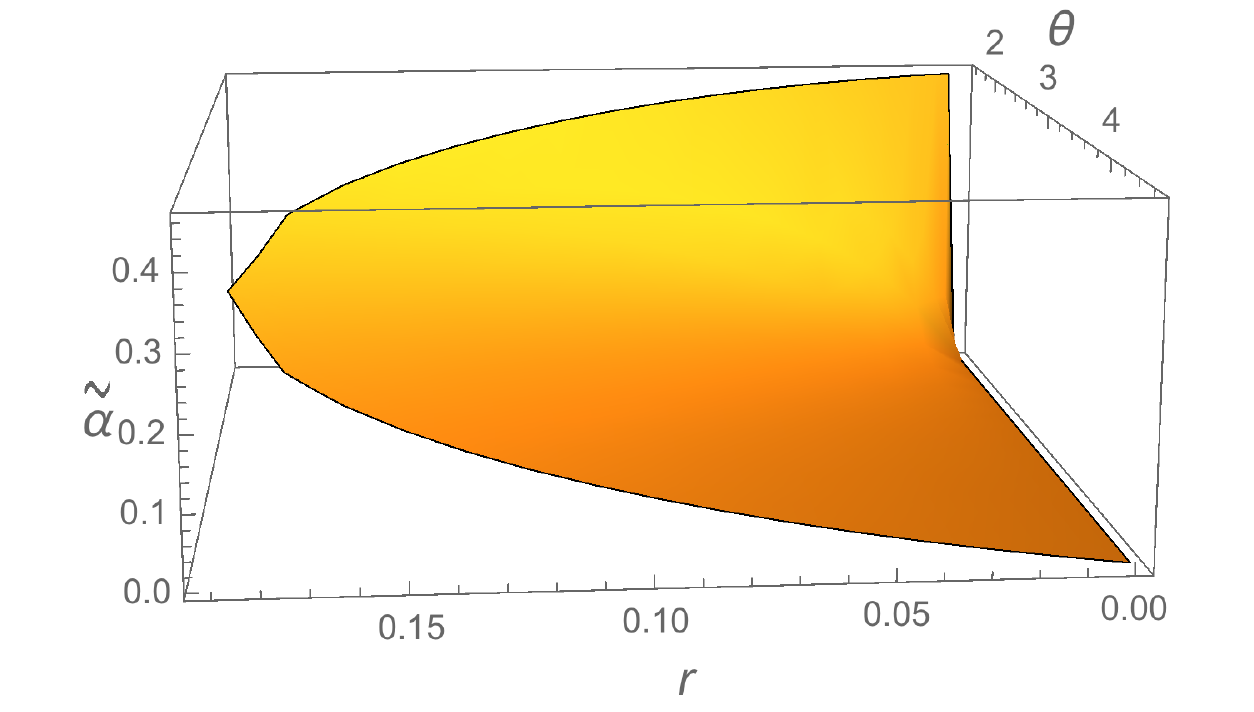}}
\caption{Plot of $\tilde\alpha= -\frac{r(1+\sin\theta)}{\cos\theta}$ against $r$ and $\theta$ in the domain $r<r_{\text{max}}$ and $\theta\in(\theta_1,\theta_2)$.}
\label{alpha tilde plot}
\end{figure}

 When $M_2^2<0$, i.e., $c_2<0$,  then $\tilde m_{\text{eff},1}^2<0$ and $\tilde m_{\text{eff},2}^2>0$ in the low temperature regime.  This means that in this regime the baryon symmetry in  the first sector is spontaneously broken and the baryon symmetry in  the second sector is  unbroken. On the other hand, in the high temperature limit, as we have already seen, the baryon symmetry in  the first sector is  unbroken and the baryon symmetry in  the second sector is spontaneously broken. Therefore, as one goes from the low temperature regime to the high  temperature regime, the baryon symmetry in the first sector would be restored at some critical temperature $T_{c1}$ and the baryon symmetry in the second sector would be broken  at a possibly different critical temperature $T_{c2}$. At these two critical temperatures, 
 \beq
 |m_1^2|_{\mu=T_{c1}}|\sim\frac{16\pi^2}{3}\lambda T_{c1}^2\ll T_{c1}^2,\ 
  |m_2^2|_{\mu=T_{c2}}|\sim\frac{16\pi^2}{3}\lambda T_{c2}^2\ll T_{c2}^2.
 \eeq
 From the relation between the two renormalized masses (squared), we get
  \beq
 |m_2^2|_{\mu=T_{c1}}|=\frac{|m_1^2|_{\mu=T_{c1}}|}{\tilde \alpha},\ 
 |m_1^2|_{\mu=T_{c2}}|=\tilde\alpha |m_2^2|_{\mu=T_{c2}}|
 \eeq
 where $\tilde\alpha\equiv -\frac{r(1+\sin\theta)}{\cos\theta}$. Just like $\alpha$ in the previous case, $\tilde \alpha<1$ in the domain $r<r_{\text{max}}$ and $\theta \in (\theta_1,\theta_2)$ as can be seen from the plot given in figure \ref{alpha tilde plot}. Therefore,
 \beq
  |m_1^2|_{\mu =T_{c2}}|\lesssim\frac{16\pi^2}{3}\lambda T_{c2}^2\ll T_{c2}^2.
\eeq
Note that $|m_i^2|_{\mu=T_{c2}}|\ll T_{c2}^2$ means that the expressions of the thermal masses (squared) given in \eqref{thermal masses on the fixed circle} are reliable near the critical temperature $T_{c2}$.  $ |m_1^2|_{\mu =T_{c2}}|\lesssim\frac{16\pi^2}{3}\lambda T_{c2}^2$ further implies  that near the critical temperature $T_{c2}$, $m_{\text{eff},1}^2$ satisfies the condition
\beq
  |m_{\text{eff},1}^2|\sim\frac{16\pi^2}{3}\lambda T^2.
\eeq
This means that we can rely on the perturbative analysis with $\mu=T$ in this regime. 

Near the other critical temperature ($T_{c1}$) the perturbative analysis can break down if $\tilde \alpha$ is very small. To avoid this, we choose a system where $r=0.1$ and $\theta=2.1$ for which $\tilde \alpha\approx 0.369$. In this system, we have
  \beq
 |m_2^2|_{\mu=T_{c1}}|\approx 2.71|m_1^2|_{\mu=T_{c1}}|\sim \frac{16\pi^2}{3}\lambda T_{c1}^2.
 \eeq
 Therefore, in this case we can rely on the perturbative analysis even near the critical temperature $T_{c1}$. In figure \ref{effective masses: case 2, subcase 2} we provide the plots of the rescaled effective masses (squared) against $t=\ln(T/\Lambda)$ for $r=0.1$, $\theta=2.1$, $\lambda=0.001$, $c_1=0$ and $c_2=-0.1$. 
 \begin{figure}[H]
\begin{subfigure}{.5\textwidth}
  \centering
  \scalebox{1}{\includegraphics[width=.8\linewidth]{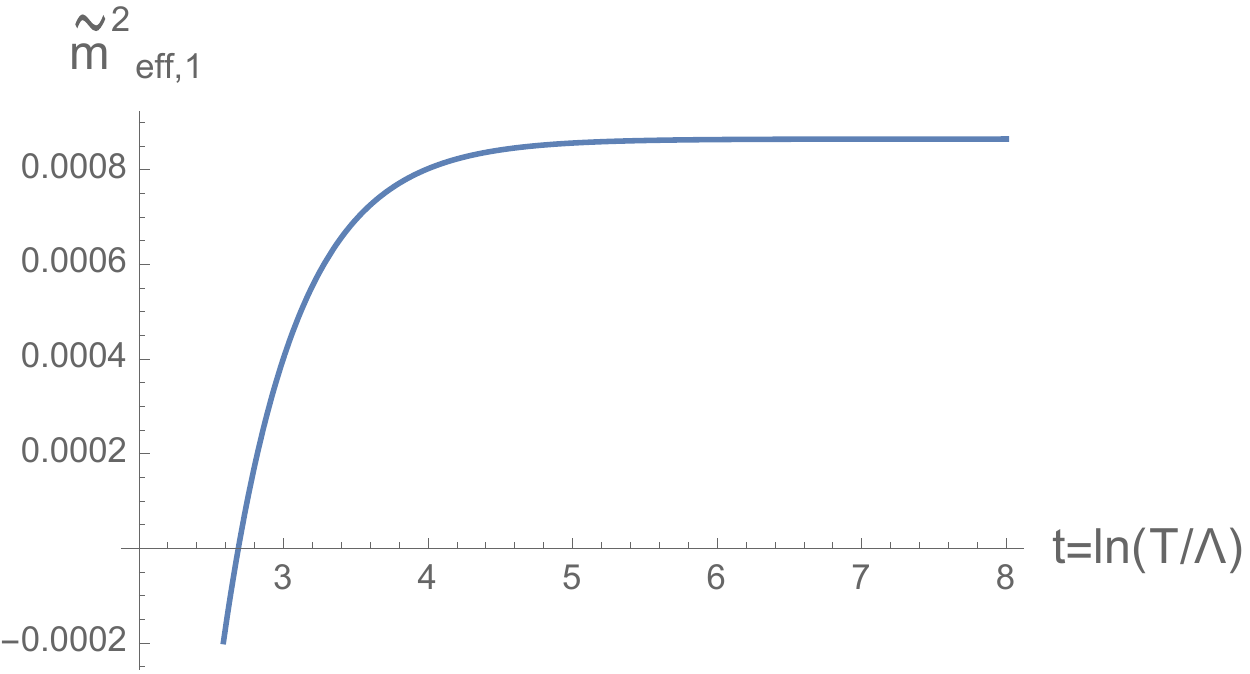}}
\end{subfigure}
\begin{subfigure}{.5\textwidth}
  \centering
  \scalebox{1}{\includegraphics[width=.8\linewidth]{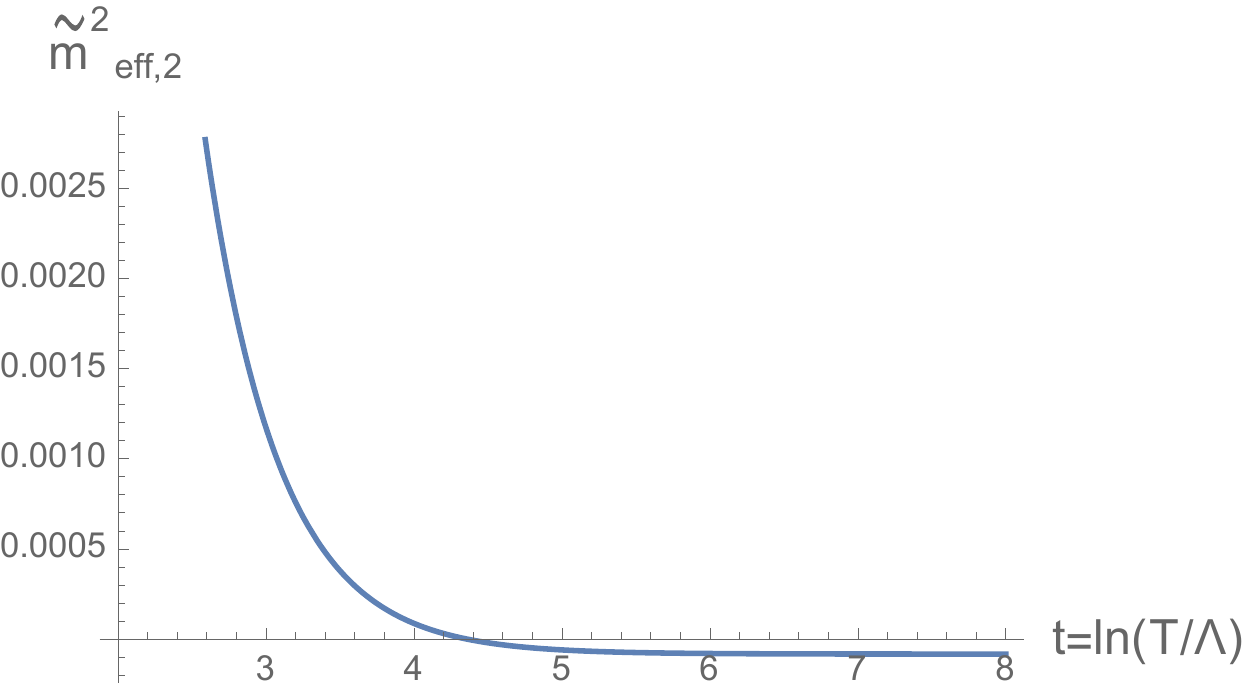}}
\end{subfigure}
\caption{Plots of the  rescaled  effective masses (squared) against $t=\ln(T/\Lambda)$  for $r=0.1$, $\theta=2.1$, $\lambda=0.001$, $c_1=0$, $c_2=-0.1$:   $\tilde m_{\text{eff},1}^2<0$ and $\tilde m_{\text{eff},2}^2>0$  in the low temperature regime. In the high temperature limit, $\tilde m_{\text{eff},1}^2$  saturates at a positive value while $\tilde m_{\text{eff},2}^2$ saturates at a negative value. This indicates that as the temperature is increased, there are two phase transitions. The first phase transition corresponds to the restoration of the baryon symmetry in the first sector. At the second phase transition, the baryon symmetry in the second sector is spontaneously broken. The two critical temperatures $T_{c1}$
 and $T_{c2}$ are given by $t_{c1}=\ln(T_{c1}/\Lambda)\approx 2.688$ and $t_{c2}=\ln(T_{c2}/\Lambda)\approx 4.355$.
}
\label{effective masses: case 2, subcase 2}
\end{figure}
As we can see from these plots, $\tilde m_{\text{eff},1}^2$ and $\tilde m_{\text{eff},2}^2$ both flip their signs as the temperature is increased. The critical temperatures ($T_{c1}$ and $T_{c2}$) corresponding to these phase transitions are given by $t_{c1}=\ln(T_{c1}/\Lambda)\approx 2.688$ and $t_{c2}=\ln(T_{c2}/\Lambda)\approx 4.355$.

Since the perturbative analysis with $\mu=T$ is   valid near the critical temperature $T_{c2}$ for different values of $(r,\theta)$ in the domain $r<r_{\text{max}}$, $\theta\in(\theta_1,\theta_2)$, we can provide a general expression for $T_{c2}$. This critical temperature can be evaluated by demanding that $\tilde m_{\text{eff},2}^2=0$ at $T=T_{c2}$. This gives us the following expression for $t_{c2}\equiv \ln(T_{c2}/\Lambda)$:
\begin{equation}
\begin{split}
& t_{c2}=-\frac{1}{2-8R_0}\ln\Bigg[-\frac{r}{c_2 \cos\theta}\Bigg\{R_0\Big(\frac{\cos\theta}{r}-\sin\theta\Big)+\frac{3}{4}\lambda\Bigg\}\Bigg].
\end{split}
\end{equation}
In figure \ref{tc2 : case 2, subcase 2}, we plot $t_{c2}$ against $\theta$ in the domain $\theta\in (\theta_1,\theta_2)$ for $r=0.1$, $\lambda=0.001$, $c_1=0$ and $c_2=-0.1$.
\begin{figure}[H]
  \centering
  \scalebox{0.6}{\includegraphics[width=.8\linewidth]{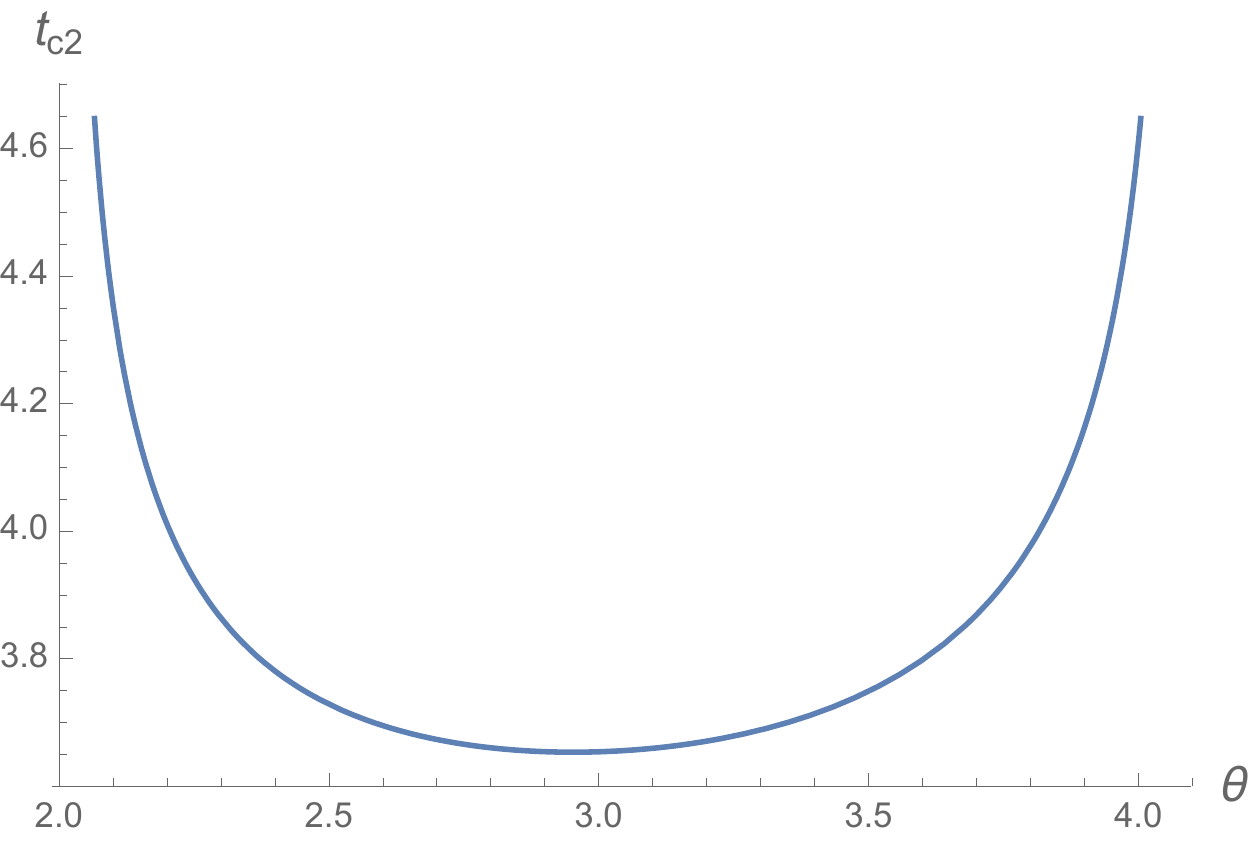}}
\caption{Plot of $t_{c2}$  against $\theta$  for $r=0.1,\ \lambda=0.001, \ c_1=0,\ c_2=-0.1$:  $t_{c2}\rightarrow\infty$ as $\theta\rightarrow\theta_{1}$ or $\theta\rightarrow\theta_{2}$.}
\label{tc2 : case 2, subcase 2}
\end{figure}
From this plot, we can see that $t_{c2}$ increases sharply as one approaches the edges of the domain $(\theta_1,\theta_2)$, and $t_{c2}\rightarrow\infty$ as $\theta\rightarrow\theta_1$ or $\theta\rightarrow\theta_2$. This means that at these edges, the baryon symmetries in  the second sector  remains unbroken at all temperatures where the perturbative analysis is valid.

\subsection{Summary of the results and comments on the case $M_1^2\neq 0,\ M_2^2\neq 0$}
\label{subsec:Massive case summary}

In this section we studied the systems that are obtained by introducing masses to the scalar fields in the RDB model. We restricted our attention to the systems that flow to the UV fixed points  exhibiting thermal order in the second sector. Due to this behavior of the systems in the UV regime, these systems remain in a persistently symmetry-broken phase in the high temperature limit. We found that as the temperature is lowered, the effects of the renormalized masses of the scalar fields start becoming significant. At sufficiently low temperatures, these renormalized  masses can lead to phase transitions in the system. We studied such phase transitions for different values of  two mass$^2$ scales $M_1^2$ and $M_2^2$ that appear in the renormalized masses (squared). When both these scales are nonzero,  they determine the asymptotic behavior of the renormalized masses  (squared) in the IR and UV regimes respectively. However, to simplify the analysis, we considered the following two cases where only one of these scales is nonzero: 
\begin{enumerate}
\item $M_1^2\neq 0, M_2^2=0$,
\item $M_1^2=0, M_2^2\neq 0$.
\end{enumerate}
We showed that in each of these cases, the sign of the nonzero $M_i^2$  determines the phase transitions occurring in the system. We showed that in certain parameter regimes, one can study these phase transitions using perturbation theory with the renormalization scale $\mu=T$. In these regimes, the perturbation theory remains valid slightly below the critical temperatures corresponding to the phase transitions. We called the phases at these scales the `low temperature' phases of the systems.  These low temperature phases and the associated phase transitions for the different cases are summarized in table \ref{table: summary of the phases}. 
\begin{table}
\centering
\caption{Phase transitions for the systems in which the baryon symmetry in the first sector is unbroken and the baryon symmetry in the second sector is broken at the high temperature limit.}
\begin{tabular}{ |c| c| c| c| }
\hline
& \multicolumn{2}{|c|}{Fate of the baryon symmetry} & Phase transitions as the\\
$M_1^2,\ M_2^2$  &   \multicolumn{2}{|c|}{in the low temperature regime}  & temperature is increased\\
 \cline{2-3}
& First sector & Second sector &\\
\hline
 &  &  & Baryon symmetry in the \\
  $M_1^2>0,\ M_2^2=0$  & unbroken & unbroken &  second sector is broken \\
 & & &   at a critical temperature.\\
 \hline
 &  &  & Baryon symmetry in the \\
  $M_1^2<0,\ M_2^2=0$  & broken & broken &  first sector is restored \\
 & & &   at a critical temperature.\\
  \hline
  $M_1^2=0,\ M_2^2>0$  & unbroken & broken & No phase transition. \\
   \hline
 &  &  & 2 phase transitions: \\
&  &  &  At a critical  temperature $T_{c1}$\\
 & & & baryon symmetry in the \\
 $M_1^2=0,\ M_2^2<0$   & broken & unbroken & first sector is restored. \\
 & & &   At another critical temperature \\
& & & $T_{c2}$, baryon symmetry in the \\
& & & second sector is broken. \\
 \hline
\end{tabular}
\label{table: summary of the phases}
\end{table}
We expect similar phase transitions even outside the aforementioned parameter regimes. However, to be sure of this, one would need to improve the perturbative analysis which lies beyond the scope of this paper. 

Finally, let us end this section by commenting on the possible phase structures when both  $M_1^2$ and $M_2^2$ are nonzero. In general, it is hard to ascertain the regime of validity of the perturbative analysis near the  critical temperatures in these cases. So, the following discussion will be mostly speculative. In the presence of both $M_1^2$ and $M_2^2$, we expect the sign of $M_1^2$ to determine the phase in the low temperature regime. This is due to the fact that as the temperature is lowered, the terms with the coefficient $c_1\equiv \frac{3M_1^2}{32\pi^2\Lambda^2}$ in the rescaled effective masses (squared) given in \eqref{rescaled effective masses: general case} grow faster than the terms with the coefficient $c_2\equiv \frac{3M_2^2}{32\pi^2\Lambda^2}$. However, at intermediate temperatures the phase can depend on $M_2^2$ as well. The situations in which $M_2^2$ can play an important role in determining the phase at such intermediate temperature scales can be guessed from the low temperature phases in the presence of a single $M_i^2$ that are  given in table \ref{table: summary of the phases}. If the signs of  $M_1^2$ and $M_2^2$ are such that individually both of them lead to the same phase in a  sector at low temperatures, then  the  phases  in that sector at different temperature regimes would be  identical to those in the presence of $M_1^2$ alone.  On the other hand, if the signs of  $M_1^2$ and $M_2^2$ are such that individually they lead to  distinct phases in a  sector at low temperatures, then  there are the following two possibilities. If there is a phase transition in that sector when only $M_1^2$ is nonzero, we do not expect anything dramatic to happen in the presence of $M_2^2$ apart from a modification of the critical temperature corresponding to this phase transition. However, if there is no phase transition in that sector when only $M_1^2$ is nonzero, then a pair of phase transitions might be induced by the presence of $M_2^2$. Whether such  a pair of phase transitions  occurs or not  would depend on the relative magnitudes of $M_1^2$ and $M_2^2$. In case  such phase transitions indeed occur in a particular sector, the phase in that sector between the two critical temperatures would be the same as the low temperature phase in the presence of $M_2^2$ alone. We do not undertake a detailed study of such double phase transitions as the perturbation theory with $\mu=T$ may not  be appropriate for analyzing the phases near the two critical temperatures.

\begin{figure}[H]
\begin{subfigure}{.5\textwidth}
  \centering
  \scalebox{1}{\includegraphics[width=.8\linewidth]{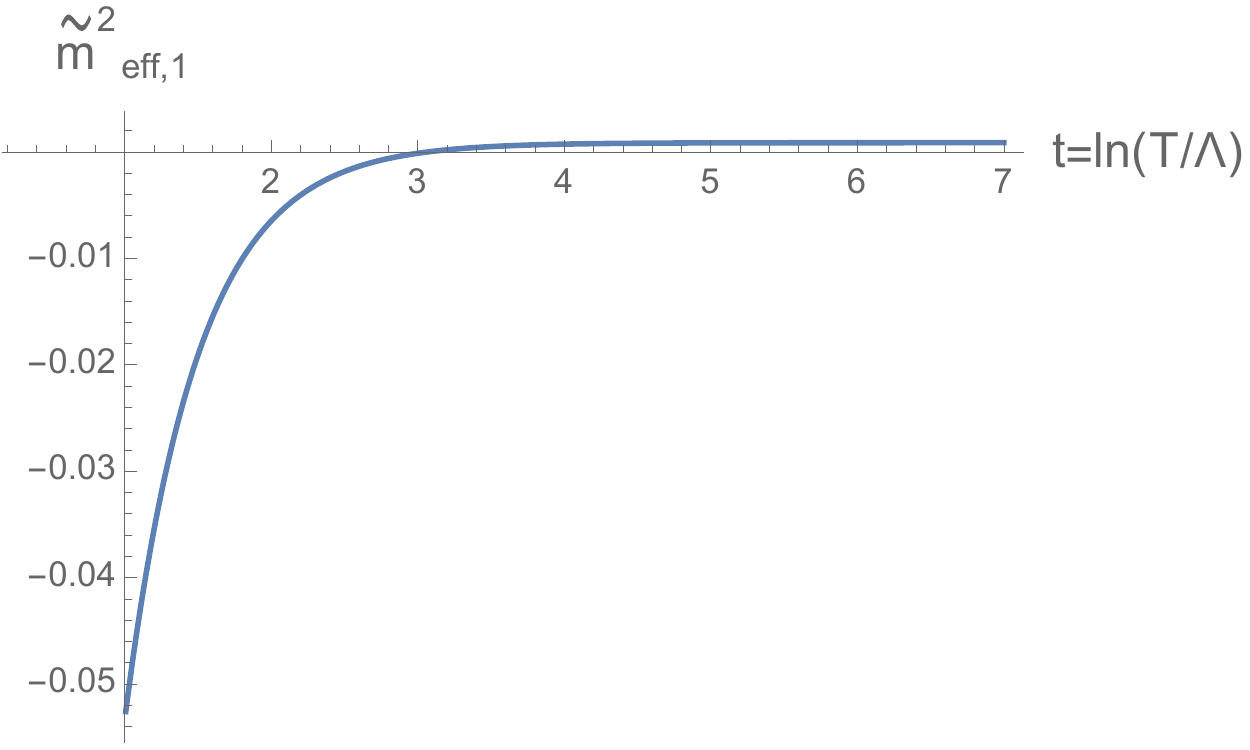}}
\end{subfigure}
\begin{subfigure}{.5\textwidth}
  \centering
  \scalebox{1}{\includegraphics[width=.8\linewidth]{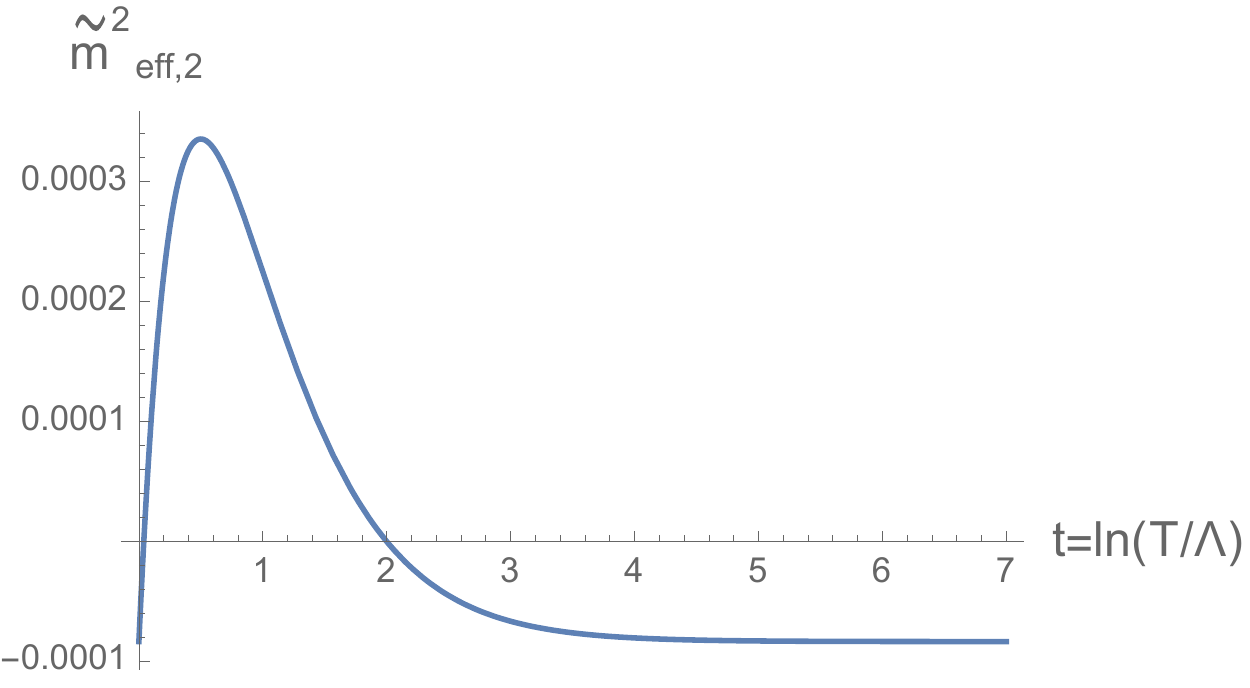}}
\end{subfigure}
\caption{Plots of the  rescaled  effective masses (squared) against $t=\ln(T/\Lambda)$  for $r=0.1$, $\theta=2.1$, $\lambda=0.001$,  $c_1=c_2=-0.2$:   A pair of phase transitions occurs in the second sector.
}
\label{effective masses: case 3}
\end{figure}

To illustrate  the possibility of such a pair of phase transitions and the associated problems with the perturbative analysis, let us consider a system where $r=0.1$, $\theta=2.1$, $\lambda=0.001$ and $c_1=c_2=-0.2$.  In figure \ref{effective masses: case 3} we plot the rescaled effective masses (squared) against $t=\ln(T/\Lambda)$ for this system. 
From these plots, we can see that at sufficiently low temperatures, both $\tilde m_{\text{eff},1}^2$ and $\tilde m_{\text{eff},2}^2$ are negative which indicates that the baryon symmetry in both the sectors are broken in this regime. This behavior at lower temperatures is determined by the negative sign of  $M_1^2$ (see table \ref{table: summary of the phases}). As the temperature is increased, due to the presence of a negative $M_2^2$,\footnote{See the low temperature phase for $M_2^2<0$ in table \ref{table: summary of the phases}.} $\tilde m_{\text{eff},2}^2$ becomes positive at a critical temperature indicating the restoration of the baryon symmetry in the second sector. When the temperature is increased further,  $\tilde m_{\text{eff},2}^2$ again becomes negative at a critical temperature and remains so at higher temperatures. This second phase transition is driven by the thermal mass (squared) $m_{\text{th},2}^2$ which begins to dominate over the renormalized mass (squared) $m_2^2|_{\mu=T}$ in this regime.\footnote{There would be a third phase transition as well where the baryon symmetry in the first sector is restored.} At this second critical temperature, the system re-enters into a phase where the baryon symmetry in the second sector is spontaneously broken. However, note that $|\tilde m_{\text{eff},1}^2|$ becomes much larger than $\lambda$ at temperatures higher than both the critical temperatures mentioned above, i.e., $|m_{\text{eff},1}^2|$ no longer remains $O(\frac{16\pi^2}{3}\lambda T^2)$ near these critical temperatures. This indicates that the perturbative analysis with $\mu=T$ is  unreliable in these regimes. We hope to come up with a more reliable analysis of  such double phase transitions in the future.

 \section{Conclusion and discussion}
 \label{sec: conclusion}

In this paper we considered  relevant and marginally relevant deformations of the large $N$ CFTs introduced in \cite{Chaudhuri:2020xxb} and showed that in the $N\rightarrow\infty$ limit, the resulting RG flows end at these CFTs in the UV regime. The gauge groups in these theories have the product form given in \eqref{gauge group}. This structure of the gauge group implies the existence of two distinct sectors in the models. The ranks of the gauge groups in the two sectors can be unequal. The matter fields in each sector transform in certain representations of the gauge group in that sector while remaining invariant under the gauge transformations in the other sector. In each of these sectors, there is a global $\mathbb{Z}_2$ (or $U(1)$)  baryon symmetry. It is the fate of these symmetries at different temperatures that was of interest to us in this work. We studied the RG flows of these theories to analyze whether the baryon symmetries in the two sectors are spontaneously broken or not at different temperature scales.   As we mentioned above, the UV fixed points of these RG flows are the large $N$ CFTs explored in \cite{Chaudhuri:2020xxb}. These CFTs are weakly coupled and they lie on a conformal manifold with the shape of a circle in the space of couplings. It was shown in \cite{Chaudhuri:2020xxb} that when the ratio of the ranks of the gauge groups in the two sectors is sufficiently away from 1, an angular interval on this circle of fixed points demonstrates thermal order characterized by the spontaneous breaking of the $\mathbb{Z}_2$ (or $U(1)$) global baryon symmetry in the sector with the smaller gauge group. It was also shown that this symmetry breaking is accompanied by  the Higgsing of half of the gauge bosons in the same sector at all nonzero temperatures.  From these facts, we concluded that the systems described by the RG flows  ending at these UV fixed points exhibit a symmetry-broken phase in the high temperature regime. We studied the IR limits of the RG flows corresponding to the different deformations and found interesting patterns of phase transitions.  Let us summarize the features of these phase transitions for the different deformations below.

The first class of deformations that we considered  involved variations of the double trace quartic couplings of the scalar fields from their values on the fixed circle. We showed that for each point on the fixed circle there is a marginally relevant deformation belonging to this class. The RG flows triggered by these marginally relevant deformations lead to a weakly coupled fixed point in the IR where the two sectors are decoupled. In fact, the theories remain weakly coupled throughout these RG flows. At the IR fixed point of these flows, the  baryon symmetries  in both the sectors are unbroken at any nonzero temperature. On the other hand, as mentioned earlier, the UV fixed points for some of these flows demonstrate thermal order characterized by spontaneous breaking of the baryon symmetry in the sector with the smaller gauge group. We showed that for each of these flows, there is an inverse symmetry breaking at a critical temperature where the baryon symmetry in the sector with the smaller gauge group is spontaneously broken.  We determined this critical temperature by finding the temperature scale at which the thermal mass of the scalar fields in the respective sector goes to zero.

The second class of deformations  involved adding masses to the scalar fields while keeping all the couplings in the model frozen at a fixed point which exhibits thermal order in the smaller sector. We studied the RG flows of these masses. This analysis showed that there are two mass$^2$ scales, $M_1^2$ and $M_2^2$, which determine the asymptotic behavior of the renormalized masses (squared) in the IR and the UV regimes respectively when both of them are nonzero. We found that in the high temperature limit, the effects of both these scales are suppressed compared to those of the thermal masses (squared) of the scalar fields, and the system remains in the same phase as the UV fixed point. Therefore, the baryon symmetry in the smaller sector remains persistently broken  in this regime. The baryon symmetry in the larger sector, on the other hand, is unbroken in the same regime. As the temperature is lowered,  the effects of $M_1^2$ and $M_2^2$ start becoming significant. At sufficiently low temperatures, they can induce phase transitions in the system. The nature of these phase transitions depends on the magnitudes and signs of $M_1^2$ and $M_2^2$. We did a detailed analysis of these phase transitions for the cases where only one of the $M_i^2$'s is nonzero. The results of this analysis for the systems where the second sector has the smaller gauge group are summarized in table \ref{table: summary of the phases}. We also found that when both  $M_1^2$ and $M_2^2$ are nonzero, in some cases there may be a pair of phase transitions occurring in a sector at different critical temperatures. We discussed the limitations of our perturbative analysis in studying the detailed features of such double  phase transitions and estimating the corresponding critical temperatures.

We now comment on a few subtleties in the analysis done in this work. Firstly, note that we relied on the 1-loop expressions of the beta functions and the thermal masses to determine the phases at different temperatures. We expect the higher loop corrections to refine the estimates of the critical temperatures while not changing the qualitative features of the phases in the different regimes. In this context, it is also important to remember that thermal perturbation theory is riddled with infrared problems \cite{Linde:1980ts, Gross:1980br}.\footnote{We refer the reader to \cite{Laine:2016hma} for detailed discussion on such IR problems.} In the models that we considered there are three  mass scales at any temperature $T$  where we could employ the perturbative analysis. These three mass scales are associated with the following modes:
\begin{enumerate}
\item The  ``hard" nonzero Matsubara modes of the different fields; these modes have masses of the  order of $\pi T$,
\item The  ``soft" zero Matsubara modes of the scalar fields and the longitudinal gauge bosons\footnote{In a symmetry-broken phase, a subset of the  transverse gauge bosons also get Higgsed. The zero Matubara modes of these Higgsed transverse gauge bosons also belong to the class of soft modes.}; these modes have masses of the  order of $\sqrt{\frac{16\pi^2}{3}\lambda} T$,
\item The  ``ultrasoft" zero Matsubara modes of the un-Higgsed transverse gauge bosons; these modes have masses of the  order of $\frac{16\pi^2}{3}\lambda T$.
\end{enumerate}
The different phases in our models are determined by the physics at the soft scale while the IR problems arise from the ultrasoft modes. They appear in the form of an infinite number of diagrams at a subleading order in the effective potential. This leads to breakdown of perturbation theory at this order. Usually  this non-perturbative physics is studied by employing the dimensionally reduced effective theory framework and the techniques of lattice gauge theories \cite{Laine:2016hma, Farakos:1994kx, Kajantie:1995kf, Kajantie:1996qd, Moore:2000jw, Kainulainen:2019kyp, Gould:2019qek, Niemi:2020hto}. Such studies on a variety  of models indicate that these non-perturbative contributions typically do not alter the qualitative features of the phases in different regimes.\footnote{They may lead to  corrections in the critical temperatures corresponding to the phase transitions.}  We expect a similar thing in the models considered in this paper. It may be useful to check this explicitly in the future.

Another point to which we would like to draw the reader's attention is that the UV fixed points of the RG flows studied in this paper are like the Banks-Zaks-Caswell fixed points in more familiar QFTs. Such fixed points are typically considered to be IR fixed points as small variations of the gauge couplings from their values at these fixed points usually lead to  asymptotic freedom. In our analysis, we avoided this by fixing the gauge couplings (as well as the single trace quartic couplings) at their fixed point values.  This could be done in the massive theories that we considered in section \ref{sec: Massive theory} because the RG flows of the masses  do not backreact on the flows of the gauge couplings (as well as the quartic couplings) in the $\overline{\text{MS}}$ scheme.\footnote{Freezing the gauge couplings and the quartic couplings  in the $\overline{\text{MS}}$ scheme at their fixed point values while allowing the masses to flow involves a fine-tuning which may take a different form in  mass-dependent renormalization schemes.} In case of the massless theories considered in section \ref{sec: marginally relevant deformations:phase transitions}, the single trace couplings could be kept frozen at their fixed point values as their beta functions are independent of the double trace couplings at the planar limit \cite{Chaudhuri:2020xxb}. 

Finally,  let us end with some comments on how finite $N$  corrections may alter the features that we have presented at the planar limit. Currently, it is not  clear to us  whether the UV fixed  points lying on the conformal manifold survive under such corrections. We refer the reader to the analysis in appendix H of \cite{Chaudhuri:2020xxb} for some perturbative computations of $1/N$ corrections which suggest that these fixed points may survive under such corrections.  Another potential problem would arise at finite N due to the fact that the double trace couplings start affecting the RG flows of the single trace couplings. This can eventually drive  these couplings away from their planar fixed point  values at sufficiently high temperatures where the cumulative effects of $1/N$ corrections become significant.  Resolving both the issues mentioned above is  complicated by the fact that to take into account $1/N$ corrections while staying in a regime where $1/N\ll$ 't Hooft couplings, one would need to re-sum the perturbative expansions in the 't Hooft couplings. It would be interesting to explore whether the thermal order persists under  $1/N$ corrections after such re-summations. We would like to return to these issues in the future.

\acknowledgments

We thank Changha Choi for useful discussions. We are also grateful to Changha Choi,  Zohar Komargodski and Michael Smolkin for their comments on an initial draft of the paper. This work is partially supported by the Israel Science Foundation Center of Excellence (Grant No. 2289/ 18).

\appendix

\section{Positivity of the  quartic terms in the thermal effective potential at leading order}
\label{app: positivity of quartic terms}
In this appendix we will show that the tree-level quartic terms in the thermal effective potential of the scalar fields remain positive throughout the RG flows studied in section \ref{subsec: RG flow}.\footnote{Our analysis will closely follow the derivation of the same result for the UV CFTs in \cite{Chaudhuri:2020xxb}.} These quartic terms have the following form:
\beq
V_{\text{quartic}}=16\pi^2\Bigg[\sum_{i=1}^2 \frac{h_i}{N_{ci}}\text{Tr}[\rho_i^2]+\sum_{i=1}^2 \frac{f_i}{N_{ci}^2}\Big(\text{Tr}[\rho_i]\Big)^2+\frac{2\zeta}{N_{c1} N_{c2}}\text{Tr}[\rho_1]\text{Tr}[\rho_2]\Bigg],
\eeq
where $\rho_i\equiv\Phi_i^T\Phi_i$. Note that  $\rho_1$ and $\rho_2$ are positive-definite matrices. From this one can easily see that a sufficient condition for the positivity of $V_{\text{quartic}}$ is
\beq
h_i>0,\ f_i>0,\ f_1 f_2-\zeta^2>0.
\label{quartic potential positivity : sufficient conditions}
\eeq
Let us now check whether these inequalities are satisfied throughout the above-mentioned RG flows. Firstly, let us note that  $h_1$ and $h_2$ are frozen at the following common value in these RG flows:
\beq
h_1=h_2=h=\Big(\frac{3-\sqrt{6}}{16}\Big)\lambda.
\eeq
Since  $\lambda$ and the coefficient $\Big(\frac{3-\sqrt{6}}{16}\Big)$ are both positive, we can see that 
$h>0$. Now, let us look at the double trace couplings $f_1=f_p+R\sin\theta$ and $f_2=f_p-R\sin\theta$. The values of $f_p$ and $R$ along these  RG flows are as follows.
\begin{equation}
\begin{split}
&f_p=f_{0p}+\frac{R_0}{2}\Bigg[\frac{(1-k)\Big(1-\tanh(8R_0 t)\Big)}{1+k\tanh(8R_0 t)}\Bigg]\ ,\ R=\frac{R_0}{2}\Bigg[\frac{(1+k)\Big(1+\tanh(8R_0 t)\Big)}{1+k\tanh(8R_0 t)}\Bigg]\ .
\end{split}
\end{equation}
From these values of the couplings along the RG flow, we get
\begin{equation}
\begin{split}
& f_{1,2}=f_p\pm R\sin\theta \geq f_p- R = f_{0p}-R_0+(1-k)R_0\Bigg[\frac{1-\tanh(8R_0 t)}{1+k\tanh(8R_0 t)}\Bigg].
\end{split}
\end{equation}
Here the quantities $k$ and $\tanh(8R_0 t)$ satisfy the following inequalities:
\beq
0<k<1,\ -1<\tanh(8R_0 t)<1.
\eeq 
From this, we can infer that 
\begin{equation}
\begin{split}
& (1-k)R_0\Bigg[\frac{1-\tanh(8R_0 t)}{1+k\tanh(8R_0 t)}\Bigg]>0.
\end{split}
\end{equation}
This leads to the positivity of the couplings $f_1$ and $f_2$ as shown below:
\begin{equation}
\begin{split}
& f_{1,2}> f_{0p}-R_0=\Bigg[\frac{\sqrt{6}}{8}-\frac{\sqrt{18\sqrt{6}-39}}{16}\Bigg]\lambda\approx 0.165\lambda>0.
\end{split}
\end{equation}
Note that for $\theta=\frac{\pi}{2}$ the positivity of $f_1$ and $f_2$ implies
\begin{equation}
\begin{split}
& f_p\pm R>0.
\end{split}
\end{equation}
This also leads to the positivity of the quantity $(f_1 f_2-\zeta^2)$ as shown below:
\begin{equation}
\begin{split}
& f_1 f_2-\zeta^2=f_p^2-R^2= (f_p+ R)(f_p-R)>0.
\end{split}
\end{equation}

Therefore, we have shown that all the conditions given in \eqref{quartic potential positivity : sufficient conditions} are satisfied by the couplings throughout the RG flows discussed in section \ref{subsec: RG flow}. Therefore the corresponding tree-level quartic terms in  the thermal effective potential of the scalar fields are positive-definite.

\section{RG flows of the masses of the scalar fields}
\label{app: 1-loop beta function of masses}

In this appendix we will derive the RG flows of the masses of the scalar fields that were introduced in section \ref{sec: Massive theory}. For this we will rely on the expressions of the 1-loop beta functions of such masses in a general gauge theory given in \cite{Luo:2002ti}. To be able to use these expressions, we need to re-express the quadratic and quartic terms involving the scalar fields  in the Lagrangian of the RDB model as follows:
\beq
& \mathcal{L}_{\text{mass}}=-\frac{1}{2}\bar m^2_{a_r i_r,b_s j_s}(\Phi_r)_{a_r i_r}(\Phi_s)_{b_s j_s},\\
&\mathcal{L}_{\text{quartic}}=-\frac{1}{4!}\bar \lambda_{a_r i_r, b_s j_s,c_t k_t, d_u l_u}(\Phi_r)_{a_r i_r}(\Phi_s)_{b_s j_s}(\Phi_t)_{c_t k_t}(\Phi_u)_{d_u l_u}
\eeq
where $\bar m^2_{a_r i_r,b_s j_s}$ and $\bar \lambda_{a_r i_r, b_s j_s,c_t k_t, d_u l_u}$ are symmetric under exchange of indices. By comparing the above forms with those given in \eqref{mass term} and \eqref{lagrangian}, we can see that
\begin{equation}
\begin{split}
& \bar m^2_{a_1 i_1,b_1 j_1}\equiv m_1^2 \delta_{a_1b_1}\delta_{i_1 j_1},\ 
\bar m^2_{a_2 i_2,b_2 j_2} \equiv m_2^2 \delta_{a_2b_2}\delta_{i_2 j_2},\ 
\bar m^2_{a_1 i_1,b_2 j_2}=\bar m^2_{a_2 i_2,b_1 j_1}\equiv 0,
\end{split}
\label{msquare matrix}
\end{equation}
\begin{equation}
\begin{split}
\bar \lambda_{a_1 i_1,b_1 j_1,c_1 k_1,d_1 l_1}\equiv &4\widetilde{h}_1\Bigg[\delta_{i_1j_1}\delta_{k_1l_1}(\delta_{a_1c_1}\delta_{b_1 d_1}+\delta_{a_1d_1}\delta_{b_1c_1})+\delta_{i_1k_1}\delta_{j_1l_1}(\delta_{a_1b_1}\delta_{c_1d_1}+\delta_{a_1d_1}\delta_{b_1c_1})\\
&\quad+\delta_{i_1l_1}\delta_{j_1k_1}(\delta_{a_1b_1}\delta_{c_1d_1}+\delta_{a_1c_1}\delta_{b_1d_1})\Bigg]\\
&+8\widetilde{f}_1\Bigg[\delta_{i_1j_1}\delta_{k_1l_1}\delta_{a_1b_1}\delta_{c_1d_1}+\delta_{i_1k_1}\delta_{j_1l_1}\delta_{a_1c_1}\delta_{b_1d_1}+\delta_{i_1l_1}\delta_{j_1k_1}\delta_{a_1d_1}\delta_{b_1c_1}\Bigg],
\end{split}
\label{symmetrised couplings def 1}
\end{equation}
\begin{equation}
\begin{split}
\bar \lambda_{a_2 i_2,b_2 j_2,c_2 k_2,d_2 l_2}\equiv &4\widetilde{h}_2\Bigg[\delta_{i_2j_2}\delta_{k_2l_2}(\delta_{a_2c_2}\delta_{b_2 d_2}+\delta_{a_2d_2}\delta_{b_2c_2})+\delta_{i_2k_2}\delta_{j_2l_2}(\delta_{a_2b_2}\delta_{c_2d_2}+\delta_{a_2d_2}\delta_{b_2c_2})\\
&\quad+\delta_{i_2l_2}\delta_{j_2k_2}(\delta_{a_2b_2}\delta_{c_2d_2}+\delta_{a_2c_2}\delta_{b_2d_2})\Bigg]\\
&+8\widetilde{f}_2\Bigg[\delta_{i_2j_2}\delta_{k_2l_2}\delta_{a_2b_2}\delta_{c_2d_2}+\delta_{i_2k_2}\delta_{j_2l_2}\delta_{a_2c_2}\delta_{b_2d_2}+\delta_{i_2l_2}\delta_{j_2k_2}\delta_{a_2d_2}\delta_{b_2c_2}\Bigg],
\end{split}
\label{symmetrised couplings def 2}
\end{equation}
\begin{equation}
\begin{split}
&\bar \lambda_{a_1 i_1,b_1 j_1, c_2 k_2,d_2 l_2}=\bar \lambda_{a_1 i_1, c_2 k_2,b_1 j_1,d_2 l_2}=\bar \lambda_{a_1 i_1, c_2 k_2,d_2 l_2,b_1 j_1}\\
&=\bar \lambda_{ c_2 k_2, a_1 i_1,b_1 j_1,d_2 l_2}=\bar\lambda_{ c_2 k_2, a_1 i_1,d_2 l_2,b_1 j_1}=\bar\lambda_{ c_2 k_2, d_2 l_2,a_1 i_1,b_1 j_1}
\equiv 8\widetilde{\zeta}\delta_{a_1b_1}\delta_{i_1j_1}\delta_{c_2d_2}\delta_{k_2l_2}.
\end{split}
\label{symmetrised couplings def 3}
\end{equation}

Now, we can use the expressions given in \cite{Luo:2002ti} to get the following 1-loop beta functions of $\bar m^2_{a_r i_r,b_s j_s}$:

\begin{equation}
\begin{split}
&(4\pi)^2 \mu\frac{d }{d\mu}(\bar m^2_{a_1 i_1,b_1 j_1})=\bar m^2_{c_1 k_1,d_1 l_1}\bar\lambda_{a_1 i_1,b_1 j_1, c_1 k_1,d_1 l_1}+\bar m^2_{c_2 k_2,d_2 l_2}\bar \lambda_{a_1 i_1,b_1 j_1, c_2 k_2,d_2 l_2}\\
&\qquad\qquad\qquad\qquad\qquad-3 \sum_{r,\gamma=1}^2 g_r^2\Lambda^{S,r\gamma}_{a_1 i_1,b_1 j_1},\\
& (4\pi)^2 \mu\frac{d }{d\mu}(\bar m^2_{a_2 i_2,b_2 j_2})=\bar m^2_{c_2 k_2,d_2 l_2}\bar \lambda_{a_2 i_2,b_2 j_2, c_2 k_2,d_2 l_2}+\bar m^2_{c_1 k_1,d_1 l_1}\lambda_{a_2 i_2,b_2 j_2, c_1 k_1,d_1 l_1}\\
&\qquad\qquad\qquad\qquad\qquad-3 \sum_{r,\gamma=1}^2 g_r^2\Lambda^{S,r\gamma}_{a_2 i_2,b_2 j_2}.
\end{split}
\end{equation}
Here 
\beq
\Lambda^{S,r\gamma}_{a_1 i_1,b_1 j_1}\equiv 2 C_2^{r\gamma}(S_1) \bar m^2_{a_1 i_1,b_1 j_1},\ \Lambda^{S,r\gamma}_{a_2 i_2,b_2 j_2}\equiv 2 C_2^{r\gamma}(S_2) \bar m^2_{a_2 i_2,b_2 j_2},
\eeq
where $C_2^{r\gamma}(S_i)$ is the quadratic Casimir of the representation in which the scalar fields in the $i^{th}$ sector transform under the $\gamma^{\text{th}}$ $G_r$ in the $r^{\text{th}}$ sector.\footnote{We remind the reader that $G_r=SO(N_{cr})$ in the RDB model.}  The values of these quadratic Casimirs are as follows:

\begin{equation}
\begin{split}
C_2^{r\gamma}(S_1)=\delta_{r1}\Big(\frac{N_{cr}-1}{4}\Big),\ C_2^{r\gamma}(S_2)=\delta_{r2}\Big(\frac{N_{cr}-1}{4}\Big).
\end{split}
\end{equation}
Substituting these  quadratic Casimirs as well as the parameters $\bar m^2_{a_r i_r,b_s j_s}$ and $\bar \lambda_{a_r i_r, b_s j_s,c_t k_t, d_u l_u}$ by their values given above, and then switching to the 't Hooft couplings introduced in \eqref{dbm: 't Hooft couplings}, we get
\begin{equation}
\begin{split}
& \mu\frac{d m_1^2 }{d\mu}=8 \Bigg[h_1\Bigg(2+\frac{1}{N_{c1}}\Bigg)+f_1\Bigg(1+\frac{2}{N_{c1}^2}\Bigg)\Bigg]m_1^2+8 \zeta\frac{N_{c2}}{N_{c1}} m_2^2-3  \lambda_1 \Bigg(1-\frac{1}{N_{c1}}\Bigg) m_1^2,\\
& \mu\frac{d m_2^2 }{d\mu}=8 \Bigg[h_2\Bigg(2+\frac{1}{N_{c2}}\Bigg)+f_2\Bigg(1+\frac{2}{N_{c2}^2}\Bigg)\Bigg]m_2^2+8 \zeta\frac{N_{c1}}{N_{c2}} m_1^2-3  \lambda_2 \Bigg(1-\frac{1}{N_{c2}}\Bigg) m_2^2.
\end{split}
\end{equation}
At the planar limit ($N_{c1}, N_{c2}\rightarrow \infty$), these beta functions reduce to
\begin{equation}
\begin{split}
& \mu\frac{d m_1^2 }{d\mu}= \Big(16 h_1+8 f_1-3  \lambda_1\Big)m_1^2+8 r \zeta m_2^2,\\
& \mu\frac{d m_2^2 }{d\mu}= \Big(16 h_2+8 f_2-3  \lambda_2\Big)m_2^2+\frac{8 \zeta}{r} m_1^2,
\end{split}
\end{equation}
where $r\equiv \frac{N_{c2}}{N_{c1}}$ in the same limit.

\bibliographystyle{JHEP} 
\bibliography{NonRe}
	
\end{document}